%% file: main.tex
\documentclass[lettersize,journal]{IEEEtran}
\IEEEoverridecommandlockouts
\usepackage{cite}
\usepackage{hyperref}
\usepackage{multirow}
\usepackage{amsmath,amssymb,amsfonts}
\usepackage{graphicx}
\usepackage{textcomp}
\usepackage{xcolor}
\usepackage{comment}
\usepackage{graphicx}
\usepackage{array}
\usepackage{xcolor}
\usepackage{tabularx}

\usepackage{makecell}
\usepackage{mathtools}
\usepackage{rotating}

\usepackage{amsmath,amsfonts}
\usepackage{array}
\usepackage[caption=false,font=normalsize,labelfont=sf,textfont=sf]{subfig}
\usepackage{textcomp}
\usepackage{stfloats}
\usepackage{url}
\usepackage{verbatim}
\usepackage{graphicx}

\usepackage{algorithmicx}
\usepackage{algorithm}
\usepackage{algpseudocode}

\def\BibTeX{{\rm B\kern-.05em{\sc i\kern-.025em b}\kern-.08em
    T\kern-.1667em\lower.7ex\hbox{E}\kern-.125emX}}
\begin{document}

\title{Architectural Exploration of Application-Specific Resonant SRAM Compute-in-Memory (rCiM)}


\author{

Dhandeep Challagundla,~\IEEEmembership{Student Member,~IEEE},
Ignatius~Bezzam,~\IEEEmembership{Member,~IEEE},
~and Riadul~Islam,~\IEEEmembership{Senior Member,~IEEE}

\thanks{D. Challagundla and R Islam are with the Department 
of Computer Science and Electrical Engineering, University of Maryland, Baltimore County, 
MD 21250, USA e-mail: {riaduli@umbc.edu}.}
\thanks{I Bezzam is with the Rezonent Inc., 
1525 McCarthy Blvd, Milpitas, CA 95035, USA e-mail: {i@rezonent.us}.}

\thanks{This research was funded in part by National Science Foundation (NSF) award number: 2138253, Rezonent Inc. award number: CORP0061, and UMBC Startup Fund.}
\thanks{Copyright (c) 2022 IEEE. Personal use of this material is permitted. 
However, permission to use this material for any other purposes must be 
obtained from the IEEE by sending an email to pubs-permissions@ieee.org.}
}
\markboth{IEEE TRANSACTIONS ON VERY LARGE SCALE INTEGRATION (VLSI) SYSTEMS}
{Shell \MakeLowercase{\textit{et al.}}: ??????}

\newcommand{\fixme}[1]{{\Large FIXME:} {\bf #1}}

\maketitle

\begin{abstract}
While general-purpose computing follows Von Neumann's architecture,
the data movement between memory and processor elements dictates the processor's performance. The evolving compute-in-memory (CiM) paradigm tackles this issue by facilitating simultaneous
processing and storage within static random-access memory (SRAM) elements. Numerous design decisions taken at different levels of hierarchy affect the figure of merits (FoMs) of SRAM, such as power, performance, area, and yield. The absence of a rapid assessment mechanism for the impact of changes at different hierarchy levels on global FoMs poses a challenge to accurately evaluating innovative SRAM designs.
This paper presents an automation tool designed to optimize the
energy and latency of SRAM designs incorporating diverse implementation strategies for executing logic operations within the SRAM.
The tool structure allows easy comparison across different array
topologies and various design strategies to result in energy-efficient
implementations. Our study involves a comprehensive comparison
of over 6900+ distinct design implementation strategies for EPFL
combinational benchmark circuits on the energy-recycling resonant
compute-in-memory (rCiM) architecture designed using TSMC
28 nm technology. When provided with a combinational circuit, the tool aims to generate an energy-efficient implementation strategy
tailored to the specified input memory and latency constraints. The
tool reduces 80.9\% of energy consumption on average across all benchmarks while using the six-topology implementation compared to
baseline implementation of single-macro topology by considering the parallel processing capability of rCiM cache size ranging from 4KB to 192KB.

\end{abstract}
\input{introduction.tex}

\input{Background.tex}

\input{propodesign.tex}

\input{result.tex}
\input{conclusion.tex}

\bibliographystyle{IEEEtran}
\bibliography{main}

\begin{IEEEbiography}
[{\includegraphics[width=1in,height=1in,clip,keepaspectratio]{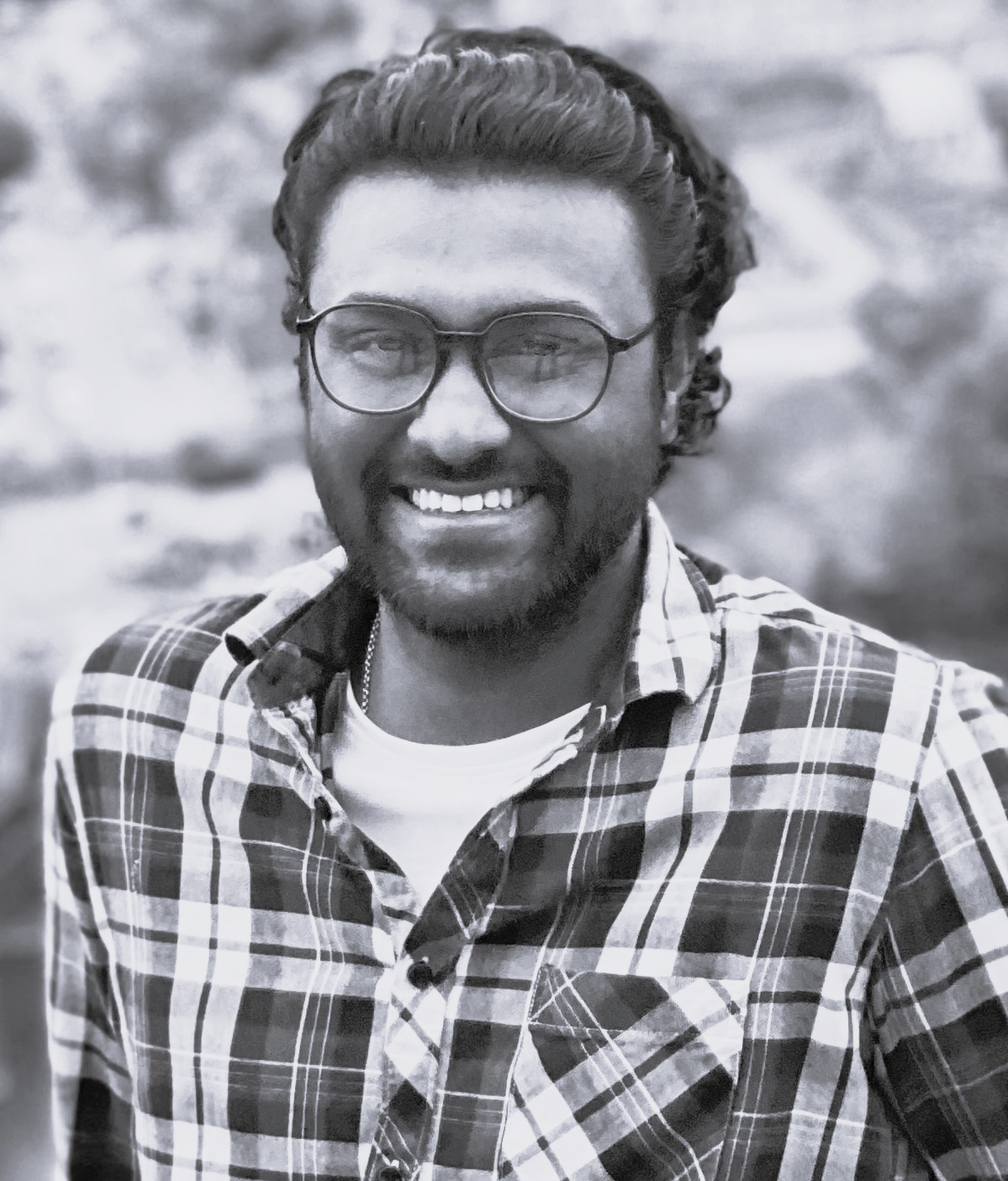}}] 
{Dhandeep Challagundla} (Student Member, IEEE) received his M.S degree from The University of Maryland Baltimore County (UMBC), MD, USA, where he is currently pursuing the Ph.D. degree with Computer Science and Electrical Engineering Department. His research interests revolve around energy-efficient computing, Compute-in-Memories, SRAM design, low-power circuit design, Mixed-signal IC design, and EDA tools.
\end{IEEEbiography}
\vspace{-2.00cm}

\begin{IEEEbiography}
[{\includegraphics[width=1in,height=1.25in,clip,keepaspectratio]{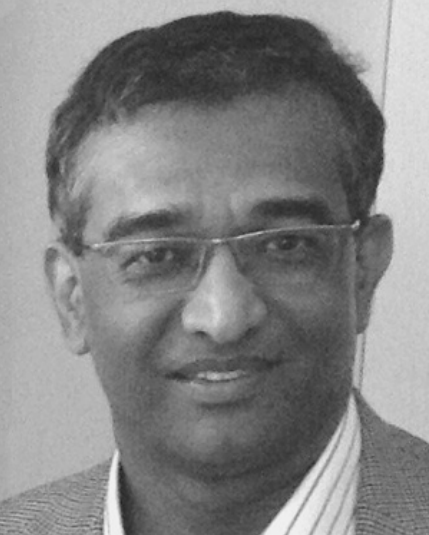}}] 
{Prof. Ignatius Bezzam} is a PhD graduate in Electrical Engineering from Santa Clara University (2015) and a Bachelor of Technology graduate of IIT Madras, India in 1983. Dr. Bezzam holds several key patents in Analog Mixed Signal Integrated Circuit (IC) design with publications in top international conferences, including the ISSCC, ESSCIRC and TCAS. Dr. Bezzam has owned 30 first silicon successes with global teams, with 33 years of next generation chip design experience in Silicon Valley, Europe and Asia.
\end{IEEEbiography}
\vspace{-1.50cm}
\begin{IEEEbiography}[{\includegraphics[width=1in,height=1.25in,clip,keepaspectratio]{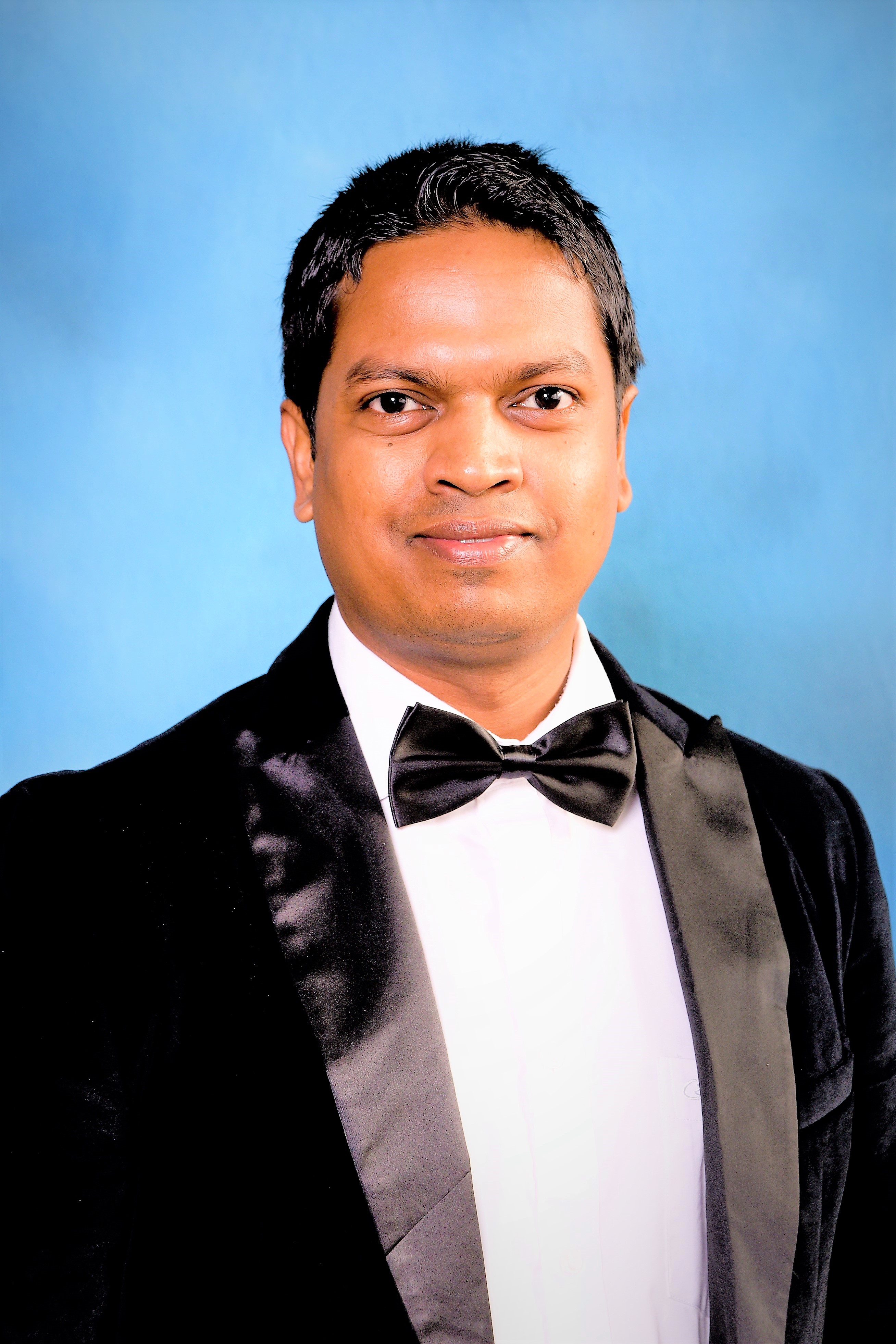}}]{Riadul Islam}
	is currently an assistant professor in the
Department of Computer Science and Electrical Engineering at the University of Maryland, Baltimore County. 
In his Ph.D. dissertation work at UCSC, Riadul
designed the first current-pulsed flip-flop/register that resulted in the 
first-ever one-to-many current-mode clock distribution networks for
high-performance microprocessors. From 2017 to 2019, he was an Assistant
Professor with the University of Michigan, Dearborn MI, USA. 
He is a senior member of the IEEE, member of the ACM, IEEE Circuits and Systems (CAS) society, the VLSI Systems and Applications Technical 
Committee (VSA-TC) of the IEEE-CAS, and IEEE Solid-State Circuits (SSC) Society. 
He holds two US patent and several IEEE/ACM/MDPI/Springer Nature journal and conference publications. 
His  current  research
interests include  digital, analog, and mixed-signal CMOS ICs/SOCs for a 
variety of applications; verification and testing techniques for analog, 
digital and mixed-signal ICs; hardware security; CAN network; CAD tools for design and analysis of
microprocessors and FPGAs; automobile electronics; and biochips. 
He is an Associate Editor of Springer Circuits, Systems and Signal Processing (CSSP) Journal.
He was a Technical Program Committee (TPC) member of the IEEE/ACM International Conference on Computer-Aided Design (ICCAD 2022), 
ACM Great Lakes Symposium on VLSI (GLSVLSI 2020, GLSVLSI 2021, GLSVLSI 2022), 57th IEEE/ACM 
Design Automation Conference (DAC) 2020 LBR Session, IEEE Computer Society Annual Symposium on VLSI (ISVLSI) 2021,  and IEEE International Conference on Consumer Electronics (ICCE) 2021.
Riadul is the
recipient of a 2021 NSF ERI award, 2021 Maryland Industrial Partnerships (MIPS) award, and 2021 Maryland Innovation Initiative (MII) award.
\end{IEEEbiography}


\end{document}

%% file: introduction.tex
\section{Introduction}
\label{sec:intro}
Cache memory remains one of the critical components in our computing system, enhancing overall performance by bridging the speed gap between the main memory (RAM) and the central processing unit (CPU). Besides, in recent years, static random access memory (SRAM)-based in-memory computing paved a promising direction to enable energy-efficient computation. However, the lack of design and automation tools to map computation on optimal SRAM architecture increases design time-to-market, resulting in higher engineering costs. This research resolves this issue by proposing an architectural exploration tool that efficiently maps logic computations to optimal cache architecture.

\begin{figure}[t!]
\begin{center}
\includegraphics[width = 0.5\textwidth]{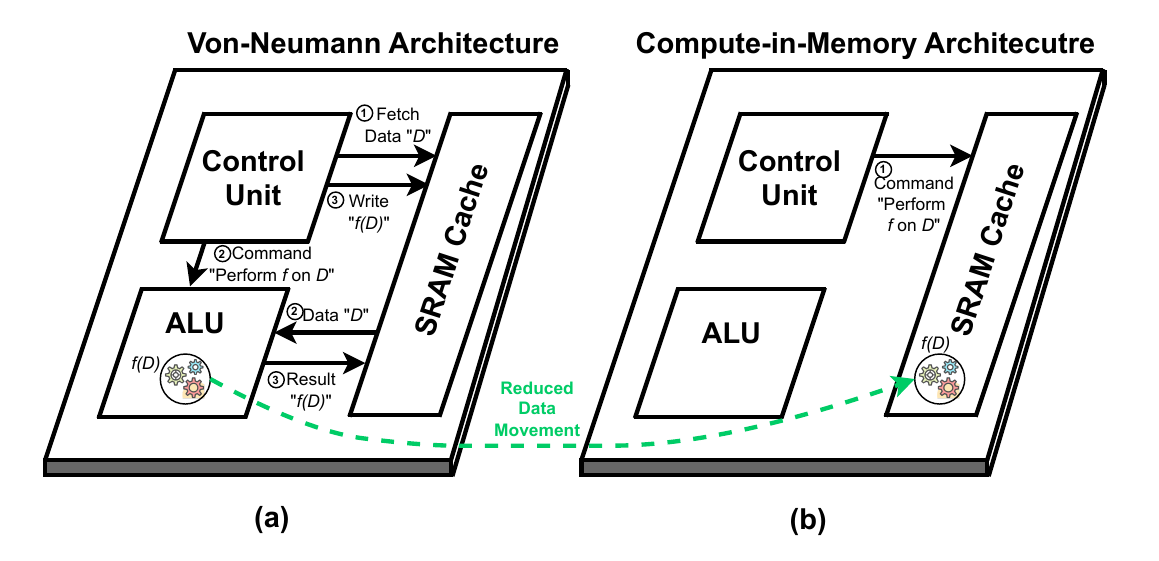}
\end{center}
\vspace{-0.50cm}
\caption{(a) Conventional Von Neumann architecture, where an operation $f$ is performed on data $D$ within the CPU, incurs high data movement overhead, which can be reduced using (b) a CiM architecture, where $f$ is computed directly within the memory, with the CPU primarily functioning as a control unit.}
\label{fig:cim_motivation}
\vspace{-0.50cm}
\end{figure}


Computing-in-memory~(CiM) architectures have emerged as highly promising solutions for data-intensive applications. They minimize data movement, enhance computational capabilities, and improve the system's overall energy efficiency by processing and storing data within cache memory. 
As shown in Figure~\ref{fig:cim_motivation}~(a), the traditional Von Neumann architecture relies on data communication between the arithmetic logic unit (ALU) and cache memory through address and data buses. However, as the CPU performance is significantly higher than the memory performance, the Von Neumann architectures often create memory bottlenecks. CiM architectures, as shown in Figure~\ref{fig:cim_motivation}~(b), mitigate the impact of large memory access latencies by performing the computations within the memory. By reducing data movement and exploiting parallelism within the memory, CiM architectures significantly enhance computational efficiency and performance.
SRAM-based CiM architectures have been heavily investigated for performing various operations, such as matrix-vector multiplication (MVM)~\cite{kroy_mvm:23,Sreekumar_mvm:24}, multiply-and-accumulate~(MAC) operations~\cite{Hechen_mac:23, si_6t:21, Gonugondla_cim:18,ali_6t:20,cheon_10t:23,Song_6t:23,biswas_10t:18,dhandeep_24_mac_cim,Swetha_mac:2023,Kavitha_cnn:24,Shengzhe_mac:24,Lu_mac:23,Wu_mac:23,Chen_mac:24}, boolean logic operations~\cite{Wang_boolean:23,Shaik_boolean:24,Zhang_cim:23,prasad_6t:23, Soundrapandiyan_8t:23, Wang_cim:19,chen_6t:21,chen_8t:21,wang_8t:20,Dinh_boolean:23,Taixin_boolean:24,Mengtian_boolean:24,Junjie_boolean:24,Hala_boolean:24}, and content-addressable memory (CAM)~\cite{Jiahao_cam:22,chen_8t:23, Wang_cim:23,jeloka_6t:16,Lin_cim:21,Yi_cam:2022} operations for fast searching operations. However, none presents a generic energy-saving architecture that spans across various applications. This work utilizes a novel series-resonance-based resonant CiM (rCiM) architecture that reduces dynamic power consumption by recycling the wasted energy during writing operations.

This work proposes an agile architectural exploration tool to map various logical operations to an optimal SRAM macro cache size. The primary objective of the tool is to facilitate the development of novel energy-efficient SRAM-based energy-recycling rCiM implementations individually designed for specific boolean logical applications.

In particular, the main contributions of the paper are as follows:
\begin{itemize}
    \item A novel resonant Compute-in-Memory (rCiM) structure that incorporates a series inductor to recycle energy dissipated during write operations.
    \item An architectural exploration toolflow that integrates open-source synthesis tools (Berkeley-ABC~\cite{Brayton_abc:2010} \& YOSYS~\cite{Yosys}) to identify the optimal SRAM configuration within a specified range of SRAM cache memory and map efficient logical operations tailored to an optimal rCiM macro size.
    \item Comprehensive analysis of 6900+ distinct logical design implementations for EPFL combinational benchmark circuits~\cite{epfl_benchmark} using 12 different SRAM topologies.
\end{itemize}

%% file: Background.tex
\section{Background}
\label{sec:background}

In recent years, considerable efforts have been dedicated to addressing the memory bottleneck associated with conventional von Neumann architectures by adopting CiM architectures. This paradigm can be implemented using both SRAM and nonvolatile memories (NVMs)~\cite{Wu:23,Duan:23,Malhotra:23,Yajuan_nvm:24,Sridharan_nvm:23,Dongre_nvm:23,Liu_nvm:24,Zhaojun_nvm:24,Choi_nvm:24}. While CiMs utilizing NVMs address static power concerns, they encounter high write energy and latency challenges. Conversely, SRAM-based CiM provides faster processing speed and robust scalability~\cite{Zhang:22}. In a recent study~\cite{Malhotra:23}, the authors propose a ferroelectric field effect transistor-based CiM technique designed for executing a single 2-operand boolean function with a single-memory access. A different study~\cite{Zhang_cim:23} achieves the implementation of an arbitrary boolean function using SRAM-based CiM. This work focuses on performing a whole combinational logic, which is crucial for SRAM-based CiMs to reduce the frequency of memory fetch operations. The diverse logical representations utilized for these combinational logic operations significantly influence the latency and overall performance of CiM architectures.

Logic synthesis takes a register transfer level (RTL) implementation, typically in Verilog or VHDL, and generates a gate-level representation of the design using a standard cell library. This work uses YOSYS synthesizer~\cite{Yosys} and ABC logic synthesizer~\cite{Brayton_abc:2010} to perform the RTL synthesis. 
The ABC takes Verilog input, and using a \textit{``strash"} function converts the input RTL into an and-inverter-graph~(AIG) graph represented as a directed acyclic graph (DAG).  
This AIG graph allows for structural optimizations to be performed~\cite{open_abcd}. This work uses four fundamental sub-graph optimizations supported by ABC, namely, \textit{``Refactor ($R_f$),"} \textit{``Rewrite ($R_w$),"} \textit{``Resubstitution ($R_s$),"} and \textit{``Balance ($B_a$)."} The $R_f$ optimization technique performs iterative collapsing and refactoring logic nodes in the AIG, aiming to reduce the AIG nodes and logic levels. Similarly, $R_w$ performs DAG-aware rewriting of the AIG network to reduce the number of logic levels. These options are significant for CiM applications, as the proposed rCiM implementation aims to perform a single level of the design hierarchy within one computational cycle. The optimization with $R_s$ is achieved by representing the logical function of a node using the existing nodes. A unique combination of these sub-graph optimizations will yield distinctive AIG implementations—the proposed algorithm in Section.~\ref{sec:proposed_algo} leverages these AIG implementations to map combinational workloads efficiently onto the rCiM architecture with diverse topologies. 

In addition, the innovative rCiM implementation employs a write driver based on series resonance and supply boosting, adopted from~\cite{Islam_sram:2021,Joshi_ibm:17,dhandeep_springer:2023,challagundla2022power,dhandeep_iscas_22,dhandeep2024system,Riadul:2018}, to significantly lower the dynamic power consumption when writing back the computational outputs. In a conventional CiM architecture, whenever a bitline discharges from a ``1" to ``0," energy gets dissipated through heat. Series LC resonance utilizes an on-chip inductor placed in the discharge path of the bitlines to store this dissipated energy and harvest it immediately into the design.

\begin{figure}[t]
\centering
\includegraphics[width = 0.5\textwidth]{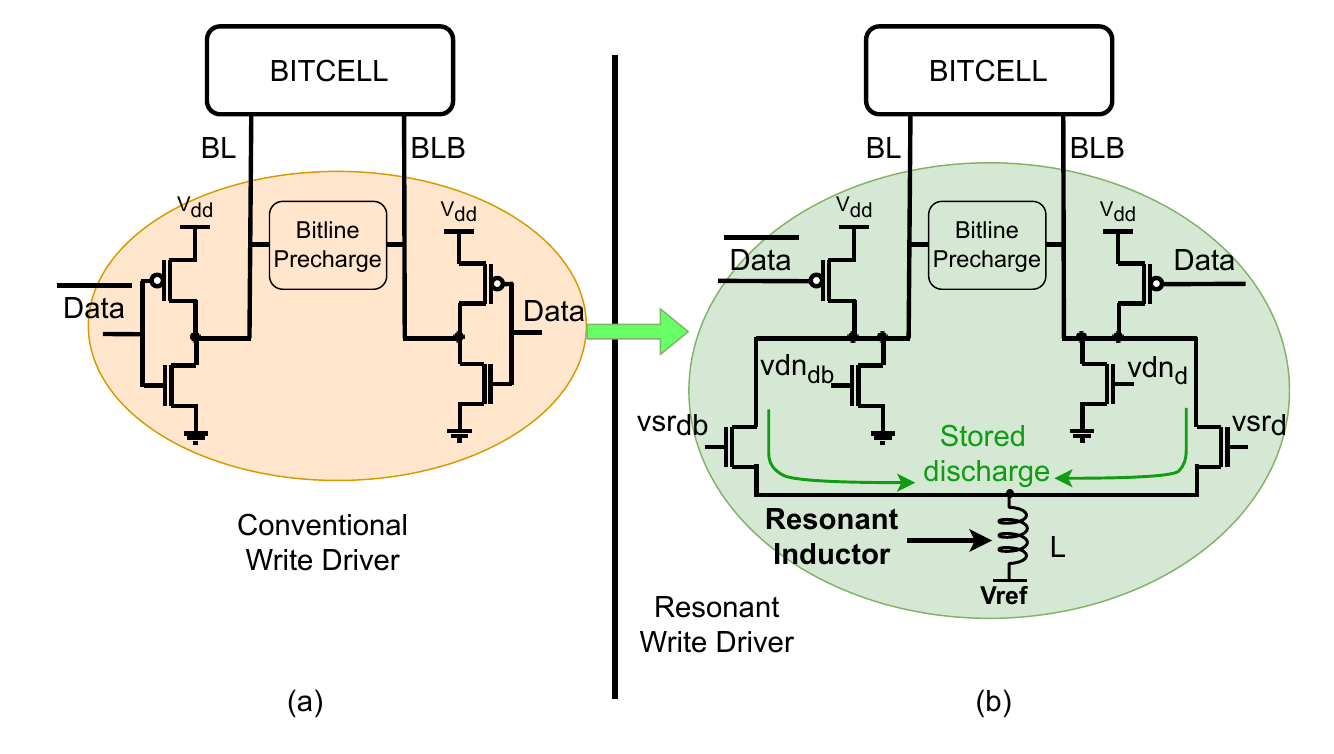}

\caption{(a) Conventional SRAM write driver exhibits high dynamic power consumption due to large bitline capacitance, whereas (b) resonant write driver recycles this dynamic power using an inductor ``$L$" placed in the discharge path, along with timed ``$vsr$" and ``$vdn$" signals.}
\label{fig:gsr_tank}

\end{figure}

Figure~\ref{fig:gsr_tank}~(a) illustrates a conventional SRAM write driver used to write data onto the SRAM using bitlines. Whenever the input data is ``0," the corresponding bitline~(BL) is driven from precharged value ($VDD$) to ground potential using the driver inverter. The resonant write driver, shown in Figure~\ref{fig:gsr_tank}~(b), employs an inductor ``$L$" to store this discharged energy. During the precharge phase, this energy is recycled back into the corresponding bitlines~(BL / BLB)~\cite{Islam_sram:2021, dhandeep_23_rcim}. At the start of the write operation, the ``$vsr$" signal is turned ``ON," enabling the inductor to store the energy discharged from bitline. Subsequently, the ``$vdn$" signal is turned ``ON" to ground the bitline fully. Once the write operation concludes and the precharge phase begins, the ``$vsr$" signal is reasserted to recycle the stored energy onto the bitline. 

Designing SRAM in scaled technologies necessitates a deep understanding of process variations, circuit dynamics, and architectural considerations. While technology scaling has facilitated the development of ever-larger cache memories, persistent challenges emerge from scaling issues. 
Open-source tools like OpenRAM~\cite{guthaus_openram:16} and VIPRO~\cite{Nalam_vipro} contribute significantly by providing essential capabilities for estimating and generating SRAM architectures but do not apply to CiM architectures as they only generate SRAM memories for read and write operations and porting for another technology is non-trivial. Recently, researchers developed OpenSAR~\cite{david_pan_open_sar:21}, a tool to design successive approximation register analog-to-digital converter (SAR ADC) based analog building blocks such as comparators and sample \& hold circuits. Another noteworthy development is AutoDCIM~\cite{chen_dcim:23}, a tool designed to generate CiM macros. These emerging tools inspire the development of an innovative architectural exploration tool that adeptly maps various logical optimizations, ensuring optimal utilization of SRAM cache architectures.

%% file: propodesign.tex
\section{Proposed Methodology}


The CiM architecture integrates a conventional SRAM cache, enabling additional computations within the same macro. This section presents a new energy-efficient CiM architecture specifically designed for performing boolean logic operations. In this paper, we proposed a novel methodology for selecting the optimal cache architecture, resulting in energy-efficient implementation tailored to a specific application.

\subsection{Proposed 10T cell}
\begin{figure}[t!]
\begin{center}
\includegraphics[width = 0.5\textwidth]{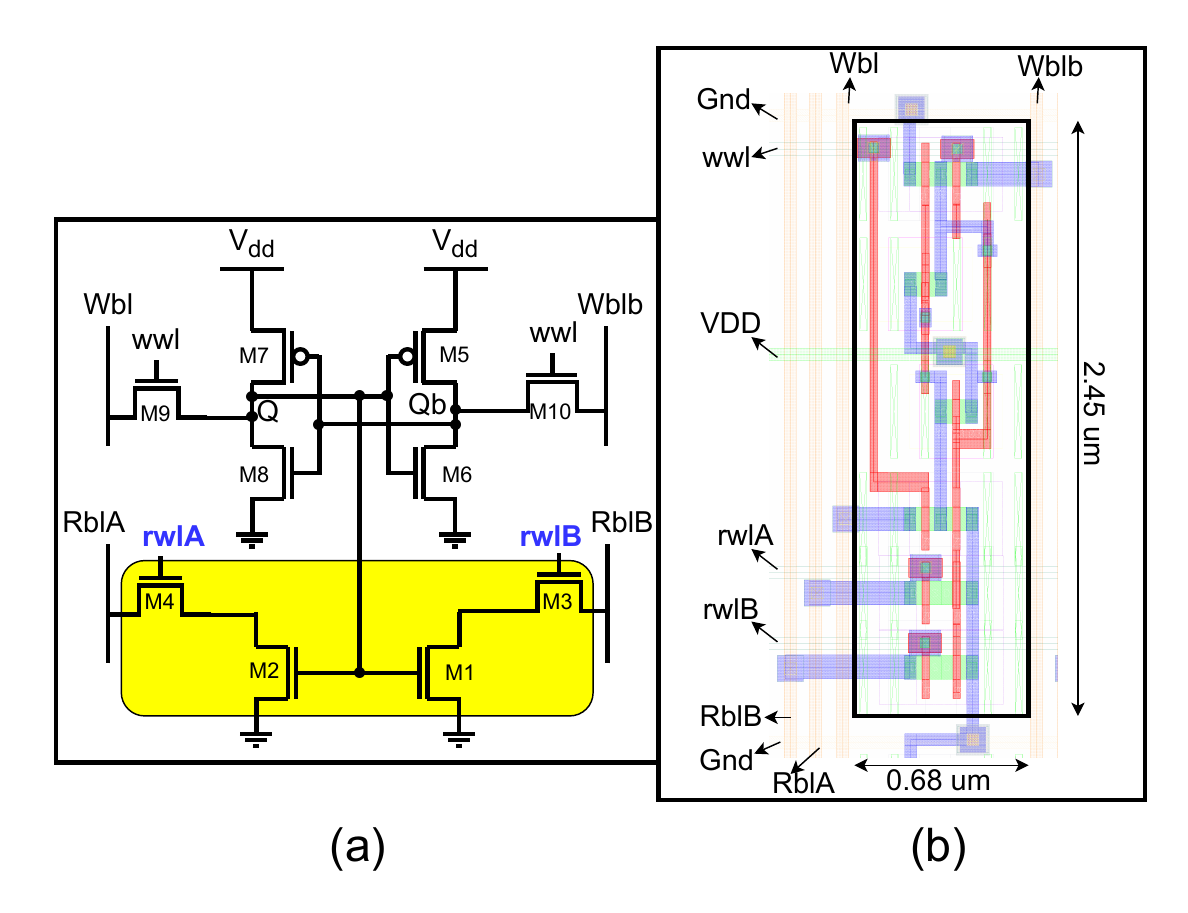}
\end{center}
\vspace{-0.50cm}
\caption{(a) The schematic of the proposed 10T SRAM cell and (b) the corresponding layout of the bitcell using up to M3 metal layers for horizontal wordlines and vertical bitlines with the area of the bitcell is $1.66\, \mu m^2$.}
\label{fig:10t_cell_layout}
\end{figure}

Figure~\ref{fig:10t_cell_layout}(a) shows the schematic of the proposed 10T SRAM cell, which builds upon a standard 6T cell architecture by incorporating four additional transistors (M1-M4). These extra transistors form a dedicated dual read-port, enhancing the cell's capability for single-bit logic operations. Figure~\ref{fig:10t_cell_layout}(b) illustrates the layout implementation of this 10T cell schematic. This layout occupies an area of $1.66\, \mu m^2$ and utilizes multiple fabrication layers, including $mpoly$ and metals. Specifically, the horizontal wordlines are routed using the M2 metal layer, and the vertical bitlines are constructed using the M3 metal layer.

\subsection{rCiM Architecture}

\begin{figure}[h]
\begin{center}
\includegraphics[width = 0.5\textwidth]{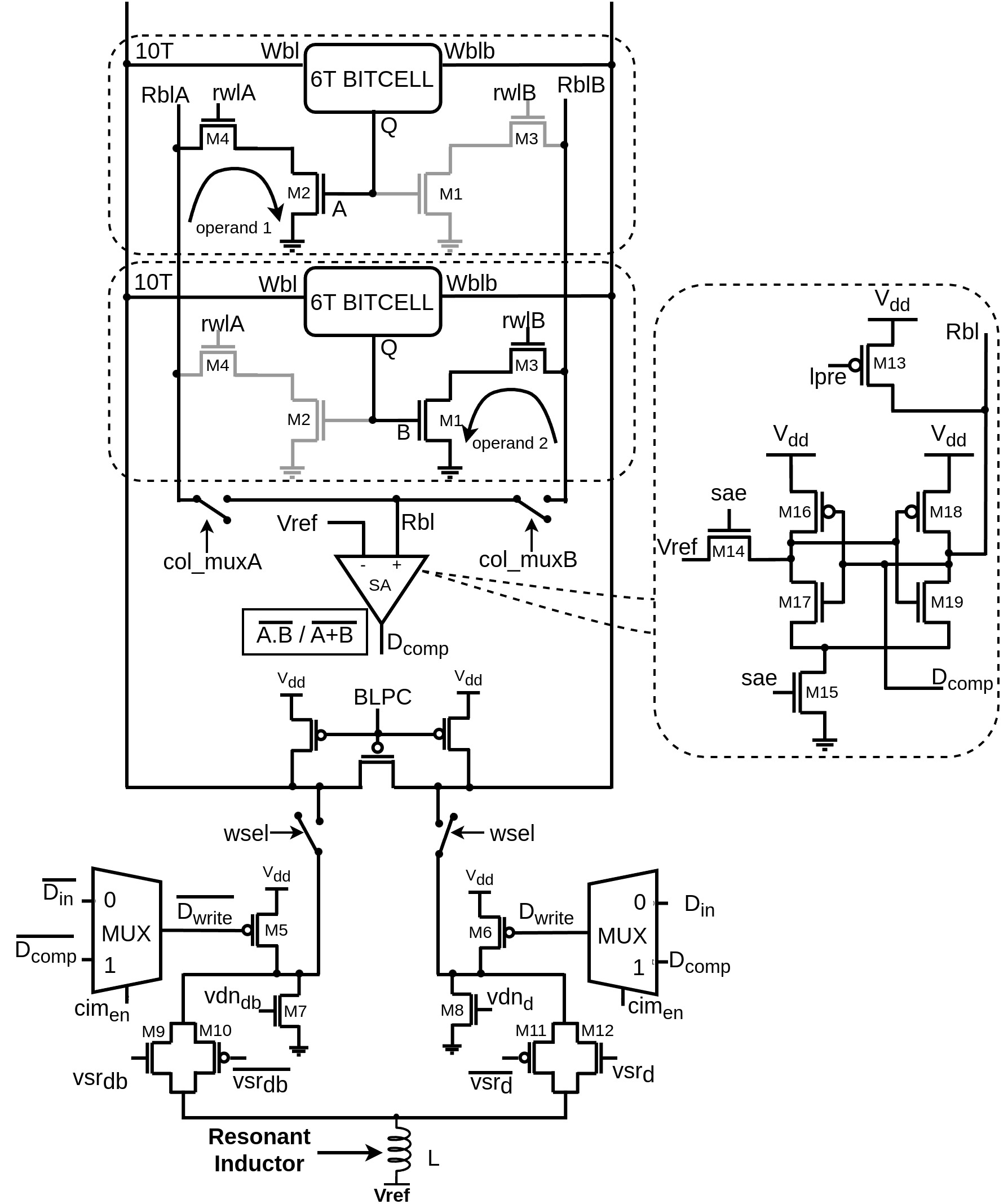}
\end{center}

\caption{10T-SRAM bitcell along with resonant write driver implementing a single logical operation using the output from the sense amplifier and writeback using an energy-recycling resonant write driver.}
\label{fig:novel_circuit}

\end{figure}

Figure~\ref{fig:novel_circuit} shows the working principle of rCIM architecture. The rCiM performs boolean logic using two 10-transistor~(10T) bit cells, as shown in Figure~\ref{fig:novel_circuit}. The transistors $M1-M4$ form a decoupled dual-read port, which allows for a large voltage swing during the conventional read operation and alleviates potential read disturb failures. Dedicated dual-read ports allow individual access to each vector operand, eliminating unidirectional computation restrictions in the SRAM array. This capability improves data retrieval efficiency, leading to enhanced system functionality and performance.

To execute a NAND2/NOR2 operation, we start by decoding the input operand addresses and simultaneously enabling the corresponding read wordlines ($rwlA~\&~rwlB$). The initially precharged read bitlines, $RblA~\&~RblB$, will eventually discharge to 0V if either of the operands corresponds to a ``1." The discharge rate of $RblA/RblB$ is dependent on whether the one-bit cell is storing a ``1" or if both the bit cells are storing a ``1." The pulse widths of read wordlines are adjusted to leverage this varying discharge rate to ensure that the $RblA/RblB$ does not completely discharge for cases``10/01" during a NAND2 operation. For a NOR2 operation, enabling the $rwlA/rwlB$ for a higher time allows the read bitlines to be driven to 0 V for cases``10/01," outputting a ``0." A programmable buffer-based pulse generator circuit is integrated with the system clock to generate the necessary $rwlA/rwlB$ pulses for performing a NAND2 or NOR2 operation. The discharge time for the NAND2 operation is approximately $150~ps$, while the NOR2 operation has a discharge time of around $350~ps$. The notable difference in discharge times contributes to the observed voltage difference between NAND2 (``01/00'') and NOR2 (``01/00'') operations, allowing for reliable distinction between these logic states. The rCiM architecture operates under a global clock frequency of 1 GHz, with all operations triggered on the rising edge of the clock. The pulse widths required for the discharge operations generated using the programmable buffer are based on the rising edges of the clock signal and a delayed clock signal. This approach ensures that the pulse width remains constant at any lower frequencies below 1 GHz, as the delay introduced by the buffer does not change.

\begin{figure}[h]
\begin{center}
\includegraphics[width = 0.42\textwidth]{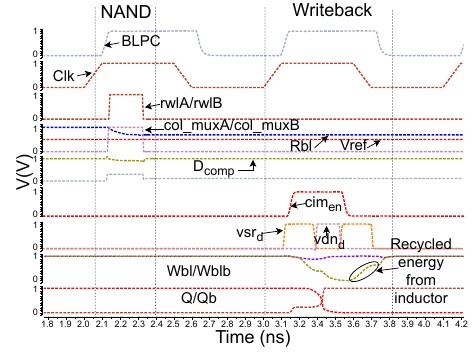}
\end{center}

\caption{The SPICE simulation confirms the correct in-memory computation considering logical NAND2 operations with ``01/10" data and a conventional energy-recycling writeback operation.}
\label{fig:functionality}

\end{figure}

Figure~\ref{fig:functionality} shows the transient simulation of performing a single NAND2 operation for cases ``10/01". The read bitlines, $RblA~\&~RblB$, are connected to one end of a single-ended sense amplifier~($SA$) through the column mux switches ($col\_muxA~\&~col\_muxB$ ) as shown in Figure~\ref{fig:novel_circuit}. The $SA$ is formed using the transistors $M13-M19$, adapted from~\cite{Rabaey:2010}. The other end of the $SA$ is connected to a reference voltage ($Vref$) which is lower than the discharge of $RblA/RblB$ during a NAND2 operation for cases ``10/01" as shown in Figure~\ref{fig:functionality}. Thus the output of the $SA$ ($D_{comp}$) will result in the output imitating a NAND2 operation by resulting a logical ``1" for all three cases (``00" \& ``10/01"). The $Vref$ signal is positioned at VDD/2 and the $rwlA/rwlB$ pulse widths are characterized such that the $Rbl$ discharge is greater than the $Vref$ voltage during the NAND2 ``10/01" cases. While performing a NOR2 operation, the $SA$ output produced a logical ``0" for all three cases (``11" \& ``10/01"). When a single vector operand is applied to both $rwlA \& rwlB$, the operation only considers two different cases (``00" \& ``11"). Thus, performing a NAND2 operation with a single vector operand results in an inversion, effectively performing a NOT operation.

The $D_{comp}$ output is latched and utilized as input data ($D_{write}$) to be written in the subsequent clock cycle by a resonant write driver. During a conventional write operation, the multiplexer selects the CPU data input ($D_{in}$). While performing CiM computations, the $cim_{en}$ signal goes high, selecting the $D_{comp}$ signal to be written into a bit cell, as shown in Figure~\ref{fig:functionality}. The energy-recycling write driver and supply-boosting, which is adapted from~\cite{Islam_sram:2021,Joshi_ibm:17}, uses a series resonant inductor to recycle the dissipated energy from write bitlines ($Wbl/Wblb$) during write operation and the precharge phase. The resonant inductor is connected to $Wbl/Wblb$ on one end, and a reference voltage ($Vref$) on the other end. To maximize the savings from the resonant inductor, the $Vref$ value is chosen to be $\frac{V_{dd}}{2}$. Whenever $Wbl/Wblb$ transitions from a logic ``1" to a logic ``0," the energy dissipated is stored in the $Vref$ node. During the precharge phase, this stored charge is emptied from the $Vref$ node, resulting in zero net currents for the whole cycle.


The resonant write driver circuit transistors $M9-M12$ shown in Figure.~\ref{fig:novel_circuit}, enable resonance by conditionally connecting the $Wbl/Wblb$ to the inductor controlled by $vsr_{d}$ and $vsr_{db}$ signals derived from the system clock. Depending upon the data, either $Wblb$ is discharged, if the input data is ``1," or $Wbl$ is discharged. For the case shown in Figure~\ref{fig:functionality}, $vsr_d$ signal is enabled to discharge the $Wblb$ signal for writing the NAND2 output of ``1" for input case ``01/10." The $vdn_{d}$ and $vdn_{db}$ signals ensure full voltage swing by completely discharging one of the write bitlines. After a successful write operation, the same transmission gates~($M11-M12$) as before are enabled to recycle the stored energy from the inductor. Hence, when the active-low bitline precharge signal ($BLPC$) is activated, there is no need to precharge the write bitlines from ``0," resulting in a decrease in the overall power consumption. The product of the bitline capacitance and the resonant inductor remains constant for a given resonant frequency. 

Utilizing a shared inductor for all the write drivers significantly minimizes the inductor's size as the bitline capacitance increases $N$ times for $N$ write drivers.

\subsection{Overall Architecture of rCiM Topologies}

\begin{figure}[b!]
\begin{center}
\includegraphics[width = 0.45\textwidth]{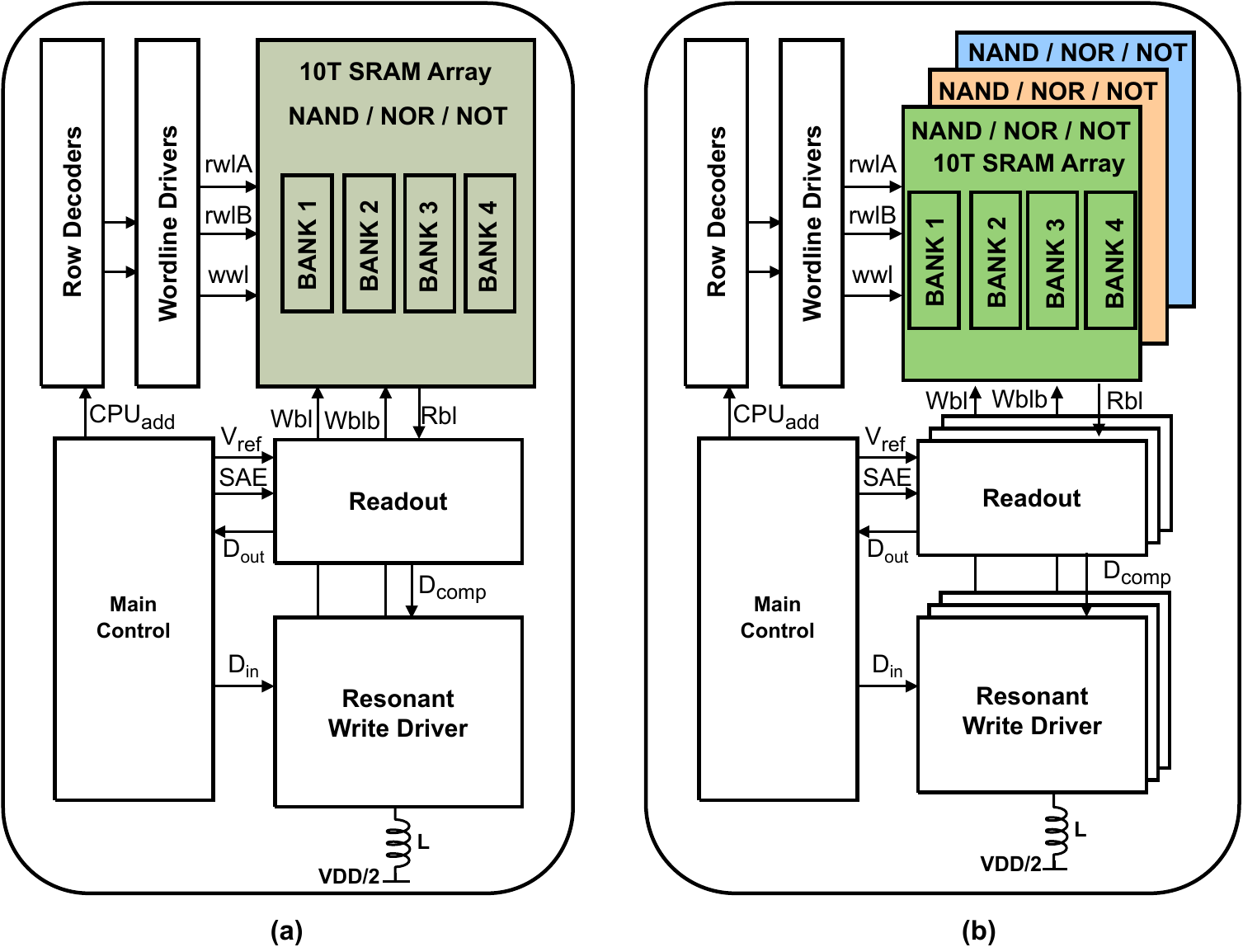}
\end{center}

\caption{Comparison of memory topology considerations for rCiM architecture, showcasing (a) single large SRAM macro or (b) multiple smaller SRAM macros.}
\label{fig:proposed_topologies}
\vspace{-0.150cm}
\end{figure}

Figure~\ref{fig:proposed_topologies} illustrates various SRAM topologies for implementing rCiM architecture. The overall architecture of rCiM includes a 10T SRAM array, a readout circuit using single-ended $SA$'s, two-row decoders enabling concurrent operands access, energy-recycling write drivers for low-power writing operations, and a central control block responsible for generating internal signals.

When considering the memory size for rCiM implementation, one can choose between a single large SRAM macro as shown in Figure~\ref{fig:proposed_topologies}~(a) or multiple smaller SRAM macros as shown in Figure~\ref{fig:proposed_topologies}~(b). The latter allows parallel execution of various logical operations, which proves beneficial for smaller designs with fewer operations in each stage, resulting in enhanced performance. However, the optimal approach for larger designs is yet to be determined—whether to increase the number of operations per stage or divide them for minimal energy consumption. The analysis in Section~\ref{sec:AIG_analysis} explains this particular aspect.

This design assigns one $SA$ for each pair of columns in the bit cell array, facilitating the execution of both conventional read operations and efficient computational processes. Consequently, the resulting architecture exhibits the capability of executing $\frac{M}{2}$ logic operations of the same kind for an SRAM bank column size of $M$.
For example, a 2KB SRAM bank with $128\times128$ SRAM bit cells can perform 64 logical operations in a single computational cycle.


Within each SRAM macro, there are several SRAM banks. By activating only one selected SRAM bank, the remaining SRAM banks enter a standby mode, resulting in a reduction in overall macro power usage. Significant dynamic power consumption in SRAM emanates from bitline charging and discharging as well as enabling wordlines. The use of multiple banks significantly contributes to the lowering of bitline power consumption.

The rCiM can be designed using two architectural configurations as depicted in Figure~\ref{fig:proposed_topologies}. The SRAM topology showcased in Figure~\ref{fig:proposed_topologies}~(a) utilizes a single macro, restricting the system to perform only one type of logical operation in a computational cycle. This architecture is particularly advantageous for scenarios with fewer logic levels but more operations within each level. Increasing the column count enables a greater number of parallel operations within a single bank, reducing the latency of the logical operation. Figure~\ref{fig:proposed_topologies}~(b) demonstrates the use of multiple SRAM macros in the rCiM. In this topology, each SRAM macro can execute a distinct logical operation. For instance, using three macros allows for the concurrent execution of NAND2, NOR2, and NOT logic operations, with each macro dedicated to one operation. This paper proposes an algorithm in Section~\ref{sec:proposed_algo} designed to choose an optimal topology from the available SRAM macro banks.

\subsection{Proposed Combinational Logic Operation Mapping Methodology}
\label{sec:proposed_algo}

\begin{figure*}[h!]
\begin{center}
\includegraphics[width = 0.8\textwidth]{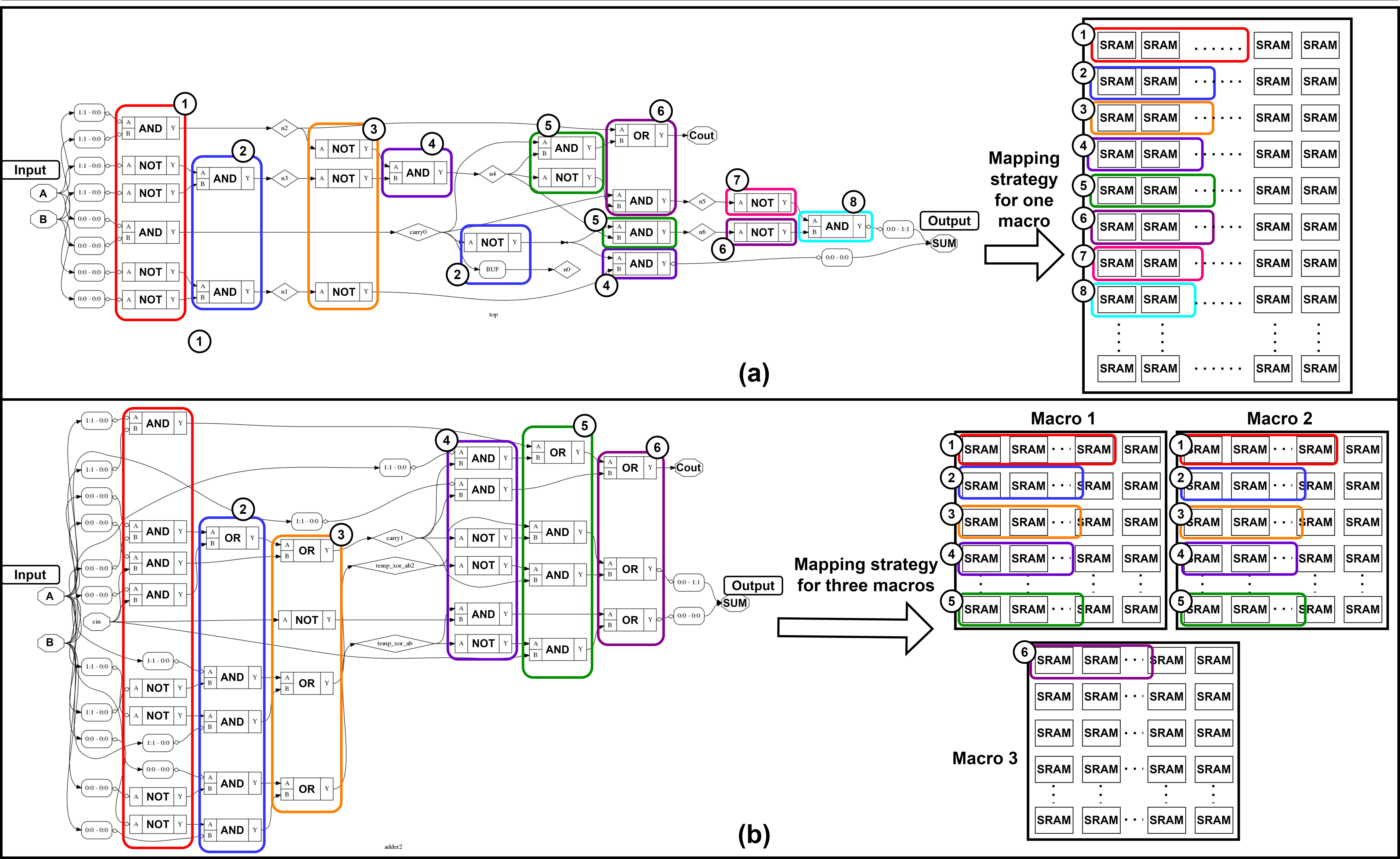}
\end{center}

\caption{The AIG graph generated using different synthesis transformations results in AIGs with different levels and different numbers of gates at each level along with mapping strategies, (a) an example AIG with eight levels mapped onto a single macro SRAM, and (b) an example AIG with six levels mapped onto a three-macro SRAM implementation.}
\label{fig:aig_graph}
\end{figure*}

Figure.~\ref{fig:aig_graph} presents two AIGs for the same 2-bit adder Verilog circuit, each generated using the ABC tool~\cite{Brayton_abc:2010} with different synthesis recipe options. These AIGs are used in YOSYS to generate netlists, which are crucial for simulating CiM designs. The variations in synthesis options result in AIGs with different levels and gate counts, significantly influencing the implementation's latency and performance in CiM.

Figure.~\ref{fig:aig_graph}(a) shows an AIG with eight levels, each level represented by a distinct color. Although it has fewer gates compared to Figure.~\ref{fig:aig_graph}(b), the higher number of levels implies greater latency when implemented in a CiM system, as each level requires one clock cycle for execution. In contrast, Figure~\ref{fig:aig_graph}(b) displays a more complex AIG in terms of gate count but with only six levels. Despite its complexity, the lower number of levels enables faster execution in CiM due to reduced clock cycles required for processing.

These diagrams effectively illustrate how different synthesis recipes affect the structure of AIGs, impacting the number of levels and the performance characteristics of the CiM systems. Thus, the choice of synthesis recipe becomes a crucial factor in optimizing computational efficiency and speed in CiM applications.
Figure~\ref{fig:aig_graph}(a) illustrates the mapping strategy for a single macro implementation, while Figure~\ref{fig:aig_graph}(b) shows the mapping strategy using a three-macro implementation. The AIG graphs are mapped in the single macro approach by assigning each logic level to a specific row or column in the SRAM array. The first level of the AIG is mapped to the first row, with its outputs stored in the second row. This pattern continues, with each level of the AIG occupying a new row and the corresponding outputs stored in subsequent rows until all AIG levels have been processed. The algorithm selects the SRAM size to ensure it can accommodate all required inputs and outputs based on the total number of gates in the design. In the three-macro implementation, the logic levels are distributed across the three macros. Each level of logic operations is divided, sorted, and assigned to a specific macro, with operands grouped accordingly. The mapping strategy then places each logic level across the SRAM rows. By aligning the data and operation execution across multiple macros, the architecture effectively manages resource constraints and maximizes throughput. If a row becomes full, the 10T bitcell allows for operands to be stored across columns as well. Since the architecture shares sense amplifiers between two columns, operands can be placed flexibly within the two columns, not strictly confined to a single row or column. This flexibility enhances the architecture’s ability to store and manage operands across multiple columns, optimizing the use of available SRAM resources.

\begin{algorithm}[h]
\renewcommand{\thealgorithm}{}
\renewcommand{\baselinestretch}{1}
\caption{\textbf{I}. Mapping combinational logic workloads to optimal resonant cache architecture}
\label{alg1:recipe_alg}
\begin{algorithmic}[1]

\State {\bf Input:} RTL netlist ($RTL$), SRAM Toplogies ($SRAM_{list}$), AIG synthesis options ($AIGsyn_{opt}$);

\State {\bf Output:} rCiM Architecture;

\State $AIG_{list} \gets CreateAIG(RTL, AIGsyn_{opt})$; \Comment{Create unique AIGs using different AIG synthesis options}~\label{alg1:create_recipe}

\ForAll{$AIG$ in $AIG_{list}$} \Comment{Loop through each AIG}~\label{alg1:for_recipe}

    

    \State $ChaAIG_{list} \gets ChaAIG(AIG)$; \Comment{Count number of hierarchy/ logic levels and logical operations in each level of the AIG}~\label{alg1:charactarize_aig}
\EndFor~\label{alg1:for_recipe_end}


\State $OptOpeAIG \gets IdentifyOptOpeAIG(ChaAIG_{list})$; \Comment{Identify AIGs with optimal number of operations}~\label{alg1:identify_optimal_operation}

\State $OptLogLevAIG \gets IdentifyOptLogAIG(ChaAIG_{list})$; \Comment{Identify AIGs with optimal number of logic levels}~\label{alg1:identify_optimal_logic_level}

\State $SRAMRange_{list} \gets IdentifySRAM(OptOpeAIG,$
$OptLogAIG, SRAM_{list})$; \Comment{Determine a set of SRAM topologies based on the total number of gate counts in the AIGs.}~\label{alg1:identify_sram_optimal}

\ForAll{$SRAM$ in $SRAMRange_{list}$} \Comment{Loop through each SRAM topology}~\label{alg1:metrics_for}
    \State $AIGMetrics_{list}[SRAM] \gets Evaluate(OptLogLevAIG,SRAM)$; \Comment{Evaluate power, latency, and energy of lowest gate count AIG for each SRAM topology}~\label{alg1:evaluate_metrics_gate}
    
    \State $AIGMetrics_{list}[SRAM] \gets Evaluate(OptOpeAIG,SRAM)$; \Comment{Evaluate power, latency, and energy of lowest logic level AIG for each SRAM topology}~\label{alg1:evaluate_metrics_level}
\EndFor~\label{alg1:metrics_for_end}


\State $BestAIG \gets FilterEnergy(AIGMetrics_{list})$; \Comment{Determine lowest energy consuming AIG}~\label{alg1:final_optimal}

\State $L_{res} \gets CalculateInductor(BestAIG.SRAM)$; \Comment{Calculate the inductor size for the chosen SRAM topology}~\label{alg1:calc_ind}
\State \textbf{Output:} rCiM Architecture $\gets$ $BestAIG.SRAM$; \Comment{Resulting rCiM architecture along with its corresponding inductor size}~\label{alg1:output_architecture}
\end{algorithmic}
\end{algorithm}

\begin{figure}[h!]
\begin{center}
\includegraphics[width = 0.5\textwidth]{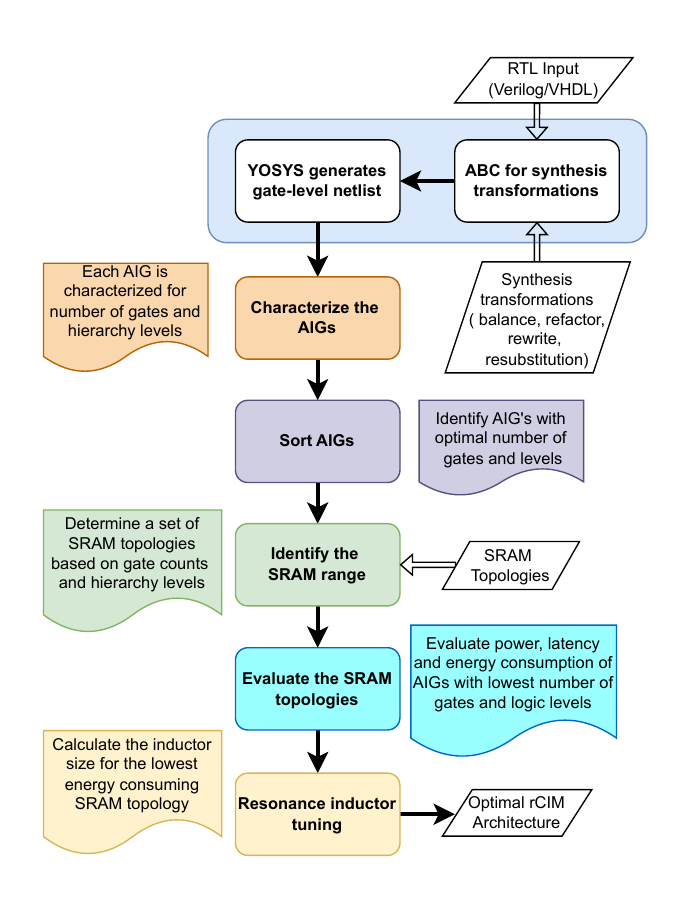}
\end{center}
\vspace{-0.50cm}
\caption{The proposed methodology flow chart shows different operations in sequential order to determine the optimal SRAM topology for a given input RTL.}
\label{fig:flow_chart}
\vspace{-0.50cm}
\end{figure}

To enable energy-efficient in-memory computation, we propose an algorithm that maps combinational logic workloads to optimal resonant cache architecture, as shown in Algorithm~I~\ref{alg1:recipe_alg}.
The algorithm takes as input the RTL netlist~(i.e., Verilog / VHDL / SystemVerilog) of the design,  AIG synthesis options ($AIGsyn_{opt}$), and the list of available SRAM toplogies~($SRAM_{list}$). The algorithm's output is an optimal energy-efficient rCIM architecture.

The algorithm starts with generating unique ($AIG_{list}$) using the AIG synthesis transformations ($AIGsyn_{opt}$) and the given RTL netlist as indicated in Line~\ref{alg1:create_recipe}. The Open-source synthesizer ABC is used to create unique AIGs using sub-graph optimizations: $B_a$, $R_f$, $R_w$, and $R_s$~\cite{Brayton_abc:2010}. The number of unique AIG synthesis transformations generated from $S$ different sub-graph optimizations is expressed by ${{\sum_{i=1}^S} {^SP_i}}$. For instance, considering $S=3$ where the provided sub-graph optimizations are $B_a$, $R_f$, $R_w$, would result in 15 unique sub-graph optimizations, such as $\{(B_a);\ (R_f);\ (R_w)\},$ $\{(B_a,R_f);\ (B_a,R_w);\ (R_f,B_a);\ (R_f,R_w);\ (R_w,B_a);\ \\ (R_w,R_f)\}$ and, $\{(B_a,R_f,R_w);\ (B_a,R_w,R_f);\ (R_f,B_a,R_w);\ \\ (R_f,R_w,B_a); (R_w,B_a,R_f);\ (R_w,R_f,B_a)\}$. This work uses four sub-graph optimizations, resulting in 64 unique AIG synthesis transformations.

When presented with an input RTL, the ABC tool initially constructs an AIG represented as a DAG. This DAG serves as the foundation for the sub-graph optimizations performing tree-balancing transformations, logic rewriting, and node reduction, which results in minimizing the delay of the design and improving logic sharing.

The flow chart in Figur.~\ref{fig:flow_chart} visually represents the proposed methodology described in Algorithm~\ref{alg1:recipe_alg}, starting with generating gate-level netlists using YOSYS and synthesis transformations using ABC. The number of gates and hierarchy levels then characterizes each AIG. These AIGs are sorted to identify those with optimal gate and logic levels. Subsequently, a set of SRAM topologies is determined based on gate counts and design cycles. The identified SRAM range is then evaluated for power, latency, and energy consumption metrics. Finally, the optimal SRAM topology is used to calculate the inductor size for the resonant inductor tuning, leading to the optimal rCiM architecture.

The For loop (Lines~\ref{alg1:for_recipe}-\ref{alg1:for_recipe_end}) iterates over every synthesized graph to characterize each AIG ($ChaAIG_{list}$). The characterization phase determines the number of stages in the design hierarchy and counts the number of logical operations at each stage.  Line~\ref{alg1:identify_optimal_operation} and Line~\ref{alg1:identify_optimal_logic_level} identifies the AIGs with optimal gate count and minimum logic level count among all the synthesized AIGs, respectively. Line~\ref{alg1:identify_sram_optimal} is used to identify a range of SRAM topologies~($SRAMRange_{list}$), considering the total number of gate counts. The range of SRAM topologies is chosen to accommodate all inputs and outputs. The memory size is chosen to be at least four times the number of gates (2 inputs + 2 outputs per gate), accounting for cases where complementary outputs are required. For example, an AIG with 128 gates requires 256 bits for inputs and 256 bits for outputs, requiring a minimum of 512 bits. Based on the AIGs chosen from Line~\ref{alg1:identify_optimal_operation} and Line~\ref{alg1:identify_optimal_logic_level}, the algorithm determines a list of suitable SRAM topologies ($SRAMRange_{list}$) from the available range of SRAM topologies.

The For loop (Lines~\ref{alg1:metrics_for}--\ref{alg1:metrics_for_end}) iterates through the library of SRAM topologies~($SRAM_{list}$) to compute the power, latency, and energy consumption metrics for the optimal SRAM ($AIGMetrics_{list}[SRAM]$) associated with optimal AIGs considering lowest gate count( Line~\ref{alg1:evaluate_metrics_gate}) and lowest logic level (Line~\ref{alg1:evaluate_metrics_level}). In lines~\ref{alg1:evaluate_metrics_gate} and \ref{alg1:evaluate_metrics_level}, power, latency, and energy metrics are derived through an analytical estimation approach combined with initial simulation data. We performed standard SRAM characterization for various topologies using post-layout analysis in Cadence Virtuoso, obtaining accurate power and latency values for different SRAM configurations. These results were used to evaluate typical read, write, precharge, and logic computation cycles for rCiM. Line~\ref{alg1:final_optimal} is used to identify optimal AIG with the lowest energy consumption among all the SRAM topologies. Line~\ref{alg1:calc_ind} uses the optimal SRAM topology to calculate the sizing of the resonant inductor. This methodology would result in the most optimal rCiM architecture implementation for the given RTL netlist.

The time complexity of the proposed methodology is determined by the number of AIGs~($n$) with $k$ levels. Additionally, the number of available SRAM topologies also plays a crucial role and is defined by $m$. The overall time complexity is expressed using BigO notation as $O(n) = O(m+n.k)$. In this work, the analysis was performed using 12 different SRAM topologies and four synthesis transformations. These four synthesis transformations resulted in 64 unique AIG synthesis options, thus setting the number of AIGs~($n$) to 64 and the size of $m$ to 12. As $m$ and $n$ are relatively small, the time complexity becomes linear and is primarily affected by the size of the levels in the AIG $k$.


%% file: result.tex
\section{Experiments and Results}
\label{sec:exp}
\subsection{Experimental Setup}

To demonstrate the efficacy of the proposed algorithm, we analyzed EPFL combinational benchmark suite circuits~\cite{epfl_benchmark} synthesized using YOSYS~\cite{Yosys}. The logic optimization of AIGs is performed using ABC~\cite{Brayton_abc:2010}. We explored 64 unique AIG synthesis options for each benchmark circuit, analyzing them across 12 different SRAM topologies for cache sizes ranging from 4KB to 192KB. The rCiM architecture was designed using TSMC 28 nm technology, and the transient simulations were performed using the Cadence Spectre simulator. Our study utilized a library of SRAM macros with sizes of 4KB, 8KB, 16KB, and 32KB. Three different topologies were employed for a comprehensive analysis of each macro size resulting in 6912 unique AIG implementations.

\subsection{AIG Transformation Analysis}
\label{sec:AIG_analysis}

\begin{figure*}[t]
\centering
\includegraphics[width = 0.88\textwidth]{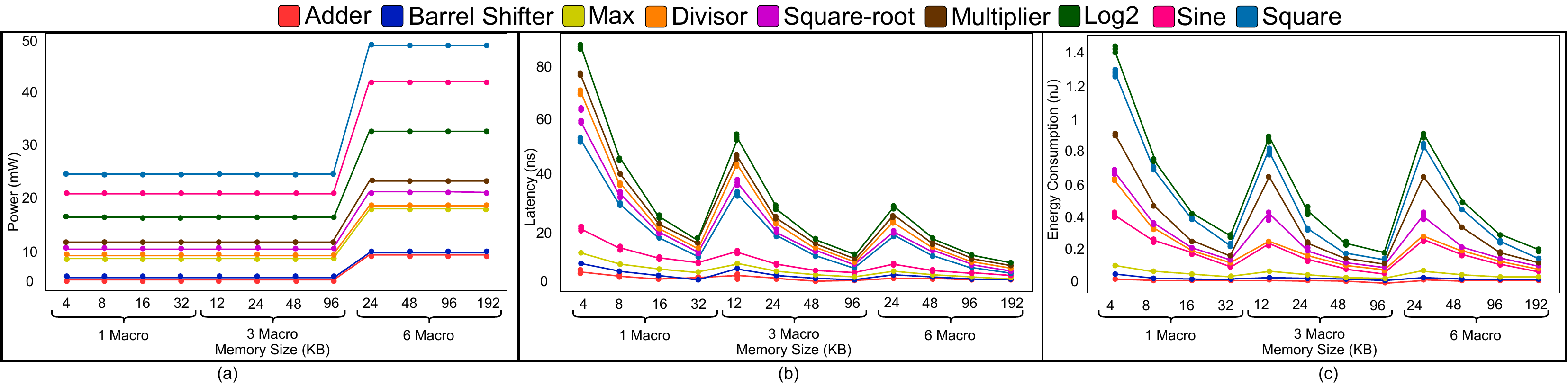}

\caption{After mapping each benchmark circuit to different SRAM architectures, we computed the power, latency, and energy; (a) power consumption remained nearly constant for single macro and 3 macro SRAMs, however, it doubled for six macro SRAMs, (b) six macro implementation achieves up to 66\% average lower latency compared single macro implementation, (c) the average energy consumption for single-macro implementations decreases up to 47\% while using an 8KB SRAM macro compared to a 4KB macro.}
\label{fig:topology_result}

\end{figure*}

Figure~\ref{fig:topology_result} compares power, latency, and energy consumption across all 6912 unique AIGs, considering 12 distinct rCiM topologies using 9 EPFL combinational benchmark circuits. The single-macro topology is limited to performing only one type of logical operation per computational cycle. In contrast, the SRAM topology with three macros can execute NAND2, NOR2, and NOT operations concurrently in each macro. For example, the three logical operations can be conducted concurrently using two macros in any six-macro implementation. 

Figure.~\ref{fig:topology_result}(a) compares the overall power consumption of each benchmark circuit. The power consumption for both the single-macro and three-macro implementations remains the same, as the total number of operations is constant. The three-macro implementation can perform three times the number of operations performed by a single-macro implementation in a single cycle, but the total number of operations required for a whole combinational logic remains the same. As a result, while the power per cycle for the three-macro implementation increases by $3\times$, it consumes $3\times$ fewer clock cycles, leading to the same overall power consumption. However, in the six-macro topology, power consumption increases by a factor of $2\times$ compared to three-macro implementation. This higher power consumption is primarily due to the doubling of power on the doubled-size macro implementation, even though the number of operations remains the same. The power per cycle for the six-macro implementation increases by $2\times$, while the number of clock cycles required to complete the operation remains the same as in the three-macro implementation, since the architecture can only perform one logic level per cycle. Thus, the total power consumption of the six-macro implementation is double that of the three-macro implementation.

Figure.~\ref{fig:topology_result}(b) depicts all the benchmark circuits' latency. In a single macro, latency decreases with an increase in macro size. On average, there is a 47\% reduction in latency when the macro area doubles from 4KB to 8KB and a 40\% reduction when the macro area goes from 16KB to 32KB. Comparatively, three-macro implementations achieve an average latency reduction of 38\%, taking advantage of the ability to perform parallel operations but incurring a $3\times$ area penalty over single-macro implementations. Similarly, six-macro implementations achieve a latency reduction of 47\% on average compared to three-macro implementations and a 66\% lower latency compared to single-macro implementations. This latency improvement results from the capability to perform more parallel operations but comes at the price of a higher area and power consumption.

Figure~\ref{fig:topology_result}(c) illustrates the energy consumption results for all benchmark circuits. The energy consumption for single-macro implementations decreases by 47\% while using an 8KB SRAM macro compared to a 4KB macro, aligning with the latency reduction as the total power consumption per benchmark computation stays nearly constant. On average, the three-macro implementations exhibit 39\% lower energy compared to single-macro implementations. Despite achieving lower latency than three-macro implementations, six-macro implementations, on average, consume 15\% higher energy due to higher power consumption.


\begin{table*}[t]
\centering
\caption{While comparing the best-case and worst-case scenarios of rCiM topologies, the three-macro implementation, with concurrent operation capabilities, demonstrates an average energy saving of 89.12\% compared to single-macro implementations with a 4KB SRAM macro size.}
\label{tab:benchmark_table}
\resizebox{1.0\textwidth}{!}{%
\begin{tabular}{|c|c|c|c|c|c|c|c|c|c|c|c|}
\hline
\textbf{Benchmark} & \textbf{Scenario} & \textbf{\makecell{ SRAM \\Macro\\Size (KB)}} & \textbf{\makecell{Macro\\Count}} & \textbf{\makecell{Synthesis\\Transformations}} & \textbf{\makecell{Level\\Count}} & \textbf{\makecell{NAND2\\Gate Count}} & \textbf{\makecell{NOR2\\Gate Count}} & \textbf{\makecell{Inverter\\Gate Count}} & \textbf{\makecell{Power\\(mW)}} & \textbf{\makecell{Latency\\(ns)}} & \textbf{\makecell{Energy\\(nJ)}} \\
\hline
\multirow{2}{*}{Adder-128} & Best-case & 16 & 3 & $R_w$, $R_f$, $B_a$ & 4 & 383 & 765 & 257 & 4.62 & 0.58 & 0.0027 \\
\cline{2-12} 
& Worst-case & 4 & 1 & $B_a$, $R_f$, $R_s$ & 4 & 170 & 1102 & 271 & 4.63 & 3.81 & 0.0176 \\
\hline
\multirow{2}{*}{Barrel Shifter} & Best-case & 32 & 3 & $R_w$, $R_f$, $B_a$ & 4 & 1474 & 1086 & 7 & 4.62 & 0.73 & 0.0034 \\
\cline{2-12} 
& Worst-case & 4 & 1 & $R_w$, $R_s$, $R_f$, $B_a$ & 4 & 1866 & 1086 & 7 & 4.63 & 6.45 & 0.0299 \\
\hline
\multirow{2}{*}{Multiplier} & Best-case & 32 & 3 & $B_a$ & 10 & 6505 & 20523 & 8638 & 11.57 & 7.395 & 0.0856 \\
\cline{2-12} 
& Worst-case & 4 & 1 & $B_a$, $R_w$, $R_s$ & 10 & 6447 & 20545 & 8639 & 11.71 & 77.06 & 0.9022 \\
\hline
\multirow{2}{*}{Sine} & Best-case & 32 & 3 & $B_a$, $R_w$, $R_s$, $R_f$ & 17 & 2341 & 4018 & 1169 & 20.80 & 2.90 & 0.0603 \\
\cline{2-12} 
& Worst-case & 4 & 1 & $R_f$, $R_s$ & 18 & 2419 & 4107 & 1120 & 20.83 & 20.09 & 0.4185 \\
\hline
\multirow{2}{*}{Max} & Best-case & 32 & 3 & $R_f$, $B_a$, $R_w$ & 8 & 655 & 2365 & 1164 & 9.25 & 1.31 & 0.0121 \\
\cline{2-12} 
& Worst-case & 4 & 1 & $R_s$, $R_f$ & 8 & 740 & 2374 & 1176 & 9.26 & 10.36 & 0.0959 \\
\hline
\multirow{2}{*}{Divisor} & Best-case & 32 & 3 & $B_a$, $R_f$, $R_s$, $R_w$ & 8 & 6696 & 18422 & 7776 & 9.26 & 6.09 & 0.0564 \\
\cline{2-12} 
& Worst-case & 4 & 1 & $R_w$ & 8 & 6828 & 18397 & 7848 & 9.39 & 70.76 & 0.6641 \\
\hline
\multirow{2}{*}{Square-root} & Best-case & 32 & 3 & $B_a$, $R_w$ & 9 & 10677 & 13561 & 6057 & 10.41 & 4.93 & 0.0513 \\
\cline{2-12} 
& Worst-case & 4 & 1 & $R_s$, $R_w$, $B_a$ & 9 & 11504 & 14621 & 4217 & 10.53 & 64.51 & 0.6792 \\
\hline
\multirow{2}{*}{Square} & Best-case & 32 & 3 & $R_w$, $R_s$, $R_f$ & 20 & 3276 & 13632 & 6308 & 24.28 & 5.66 & 0.1373 \\
\cline{2-12} 
& Worst-case & 4 & 1 & $R_f$, $R_w$, $R_s$, $B_a$ & 21 & 3131 & 13977 & 6257 & 24.36 & 53.25 & 1.2973 \\
\hline
\multirow{2}{*}{$Log_2$} & Best-case & 32 & 3 & $R_f$, $R_s$, $B_a$ & 13 & 10195 & 21848 & 7839 & 16.20 & 7.40 & 0.1198 \\
\cline{2-12} 
& Worst-case & 4 & 1 & $R_f$, $R_w$, $R_s$ & 14 & 10482 & 22348 & 7836 & 16.35 & 87.77 & 1.4351 \\
\hline
\end{tabular}%
}
\end{table*}

In Table~\ref{tab:benchmark_table}, we present a comprehensive comparison of AIG implementations for the EPFL benchmark circuits, highlighting the best and worst-case AIG implementations. Additionally, the table provides insights into the number of stages, gate counts, and synthesis transformations employed for each benchmark. The analysis uses four different synthesis options (i.e., $B_a$, $R_f$, $R_w$, and $R_s$). The analysis shows that employing multiple macros leads to the most energy-efficient design by leveraging concurrent operations. However, excessive macro use can compromise energy efficiency due to increased power consumption.

In the case of the Adder-128 benchmark, which has a small number of operations, dividing a 48KB SRAM into three macros resulted in significantly lower energy consumption. The benchmark exhibits an 85\% reduced energy consumption compared to a single 4KB macro achieved by concurrent operations. For benchmark circuits with a substantial number of operations, such as $Log_2$, employing synthesis transformations to reduce 2\% of the operations and opting for larger macros to execute a higher number of concurrent operations resulted in a 92\% reduction of energy consumption, but with a $24\times$ area penalty. In the case of the Sine circuit, with a moderate gate count, adopting a three-macro implementation of 96KB SRAM size resulted in an 85.4\% reduction in energy consumption. Similarly, using a three-macro implementation of 96KB SRAM size for the Square-root operation showcased a reduction of 93\% energy consumption compared to a single 4KB macro implementation.

In summary, this study highlights the tradeoffs between area, latency, and the SRAM topology to achieve an energy-efficient rCiM implementation. To achieve lower latency, we have two main strategies: either increase the size of a single macro or employ multiple smaller macros to carry out parallel operations. For example, in the case of the divisor benchmark circuit, the rCiM circuit achieves a latency reduction of 92\% with a $12\times$ SRAM area penalty after utilizing the three-macro SRAM topology.

\subsection{Process Variation Analysis}

\begin{figure*}[t]
\centering
\includegraphics[width = 0.88\textwidth]{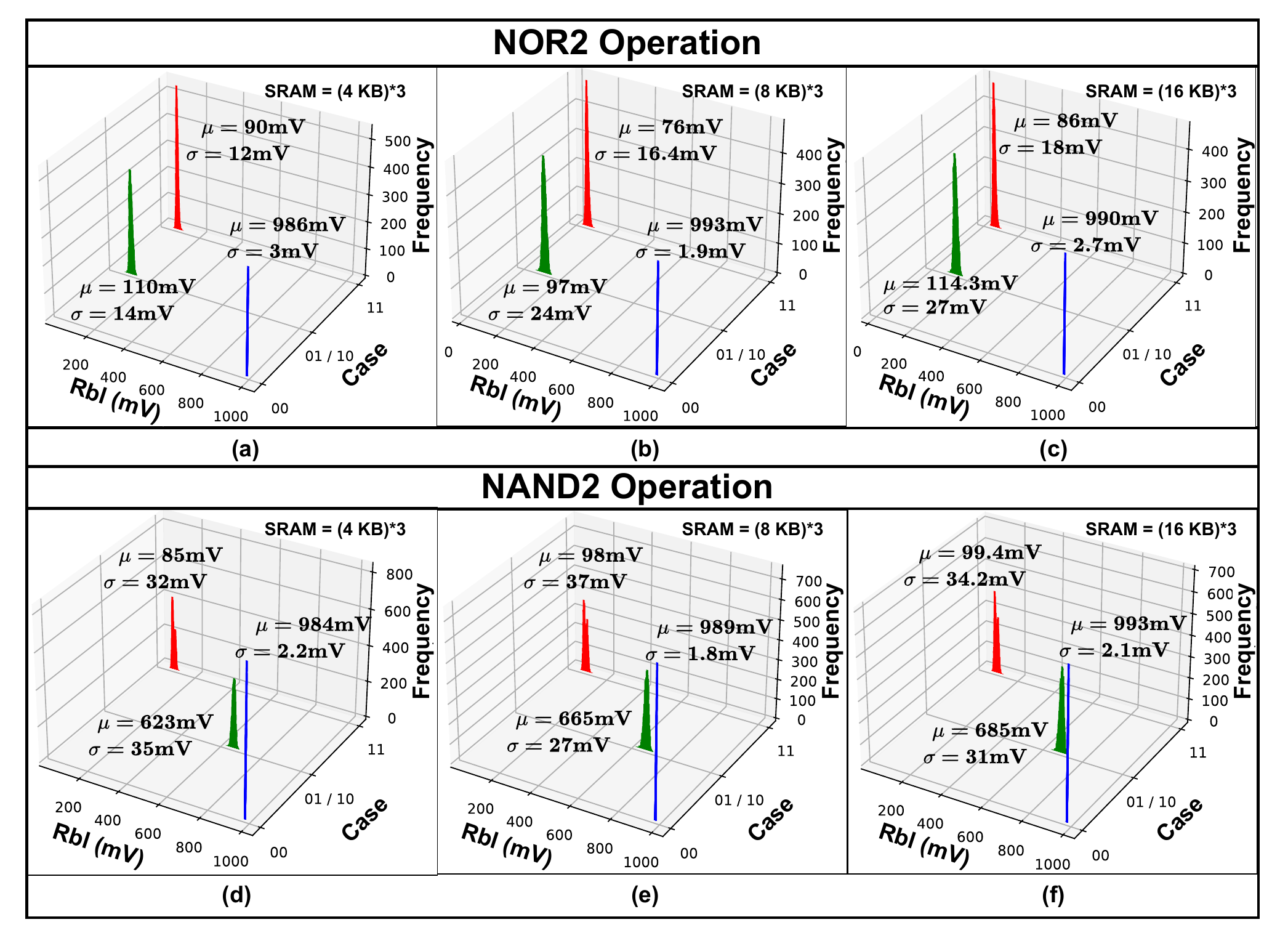}

\caption{ Monte-Carlo simulations considering 5000 samples of the $Rbl$ discharge conducted across three SRAM topologies, each under $\pm$10\% length variation with 3$\sigma$ deviations for the cases ``01/10," ``00,"  and ``11," of NAND2 and NOR2 operations.}
\label{fig:monte_carlo}

\end{figure*}

Figure~\ref{fig:monte_carlo} evaluates the robustness of the proposed rCiM architecture against process variations for all input cases. We consider three different SRAM topologies: (4 KB)$\times$3, (8 KB)$\times$3, and (16 KB)$\times$3. For each topology, 5000 samples of the $Rbl$ discharge were taken with $\pm 10\%$ length variation of all transistors under 3$\sigma$ deviations.

The NOR2 operation analysis for the three SRAM topologies is shown in Figure~\ref{fig:monte_carlo} (a), (b), and (c). For the (4 KB)$\times$3 topology shown in Figure~\ref{fig:monte_carlo} (a), the mean $Rbl$ voltages are 110~$mV$, 986~$mV$, and 90~$mV$ with standard deviations of 14~$mV$, 3~$mV$, and 12~$mV$ for cases ``01/10," ``00,"  and ``11," respectively. In the (8 KB)$\times$3 topology depicted in Figure~\ref{fig:monte_carlo}~(b), the mean $Rbl$ voltages are 97~$mV$, 993~$mV$, and 76~$mV$, with standard deviations of 24~$mV$, 1.9~$mV$, and 16.4~$mV$ for the same cases. For the (16 KB)$\times$3 topology shown in Figure~\ref{fig:monte_carlo}~(c), the mean $Rbl$ voltages are 114.3~$mV$, 990~$mV$, and 86~$mV$, with standard deviations of 27~$mV$, 2.7~$mV$, and 18~$mV$, respectively.

The NAND2 operation analysis is depicted in Figure~\ref{fig:monte_carlo}~(d), (e), and (f). For the (4 KB)$\times$3 topology in Figure~\ref{fig:monte_carlo}~(d), the mean $Rbl$ voltages for cases ``01/10," ``00," and ``11" are 623~$mV$, 984~$mV$, and 85~$mV$, with standard deviations of 35~$mV$, 2.2~$mV$, and 32~$mV$, respectively. In the (8 KB)$\times$3 topology shown in Figure~\ref{fig:monte_carlo}~(e), the mean $Rbl$ voltages are 665~$mV$, 989~$mV$, and 98~$mV$, with standard deviations of 27~$mV$, 1.8~$mV$, and 37~$mV$. Lastly, for the (16 KB)$\times$3 topology in Figure~\ref{fig:monte_carlo}~(f), the mean $Rbl$ voltages are 685~$mV$, 993~$mV$, and 99.4~$mV$, with standard deviations of 31~$mV$, 2.1~$mV$, and 34.2~$mV$, respectively.

\begin{figure*}[h]
\centering
\includegraphics[width = 0.88\textwidth]{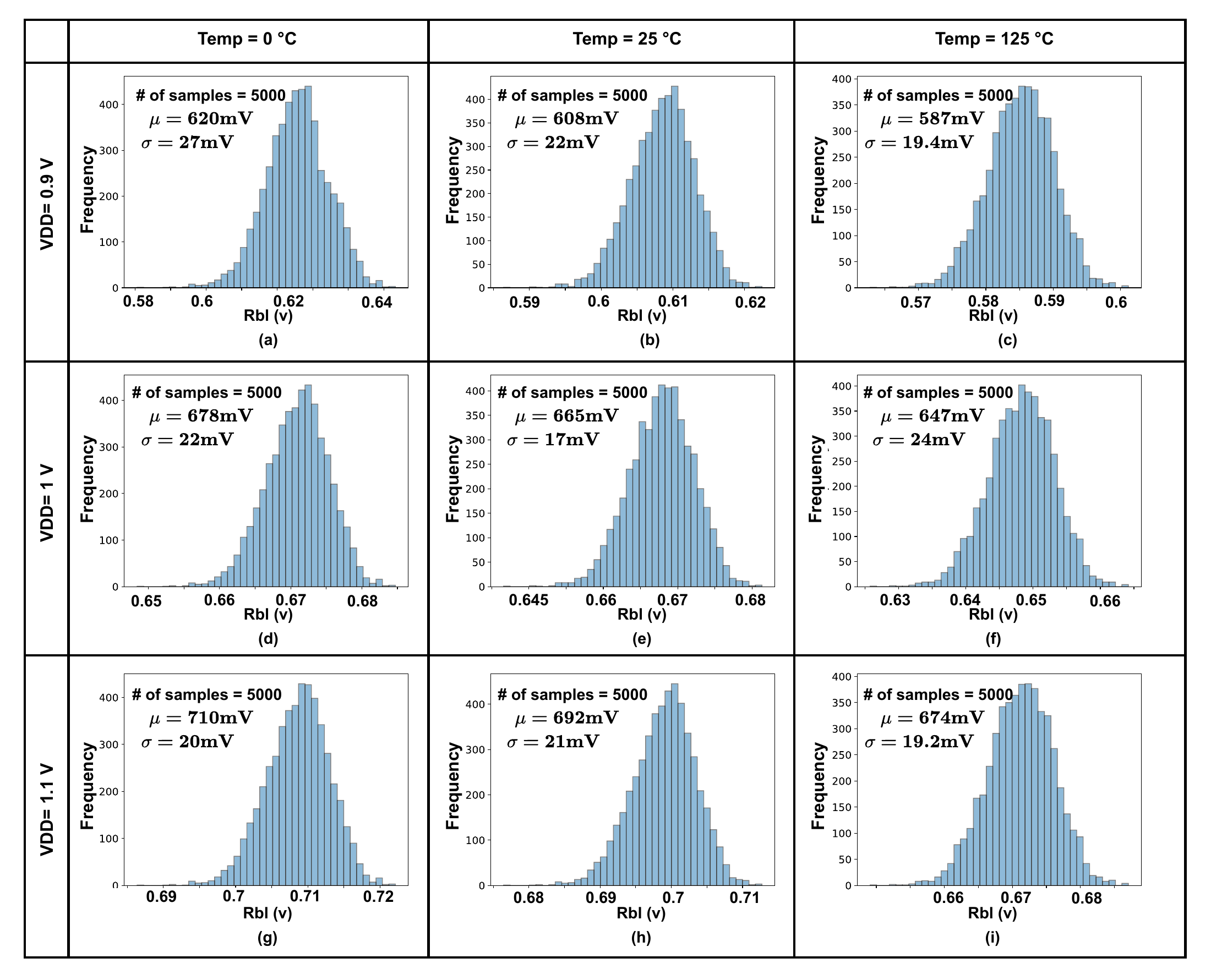}

\caption{Monte-Carlo simulations with variations in the temperature and $\pm$10\% of the supply voltage of the proposed rCiM for the borderline NAND2 01/10 input vector case considering 5000 samples under $\pm$10\% length variation.}
\label{fig:temp_voltage_monte_carlo}

\end{figure*}

Monte-Carlo simulations were performed to evaluate the impact of temperature and voltage variations on the system's performance for the borderline case ``01/10" for the (8 KB)$\times$3 SRAM topology. A total of 5000 samples were analyzed for each combination of temperature and voltage. The simulations considered three different temperatures (0°C, 25°C, and 125°C) and three voltage levels (0.9 V, 1 V, and 1.1 V). The results, depicted in Figure~\ref{fig:temp_voltage_monte_carlo}, show the $Rbl$ discharge distribution values.

At a temperature of 0°C, the $Rbl$ discharge for voltages of 0.9 V, 1 V, and 1.1 V, as illustrated in Figure~\ref {fig:temp_voltage_monte_carlo} (a), (d), and (g), respectively, are of significant importance. For 0.9 V, the mean $Rbl$ voltage is 620~$mV$ with a standard deviation of 27~$mV$. At 1 V, the mean $Rbl$ voltage is 608~$mV$ with a standard deviation of 22~$mV$. For 1.1 V, the mean $Rbl$ voltage is 587~$mV$ with a standard deviation of 19.4~$mV$.

At 25°C, the $Rbl$ discharge for voltages of 0.9 V, 1 V, and 1.1 V, as shown in Figure~\ref{fig:temp_voltage_monte_carlo}~(b), (e), and (h), respectively, have been thoroughly analyzed. The mean $Rbl$ voltage for 0.9 V is 647~$mV$ with a standard deviation of 24~$mV$. For 1V, the mean $Rbl$ voltage is 665~$mV$ with a standard deviation of 17~$mV$. For 1.1 V, the mean $Rbl$ voltage is 678~$mV$ with a standard deviation of 22~$mV$.

At a higher temperature of 125°C, the $Rbl$ discharge for voltages of 0.9V, 1 V, and 1.1 V are presented in Figure~\ref{fig:temp_voltage_monte_carlo}~(c), (f), and (i), respectively. The mean $Rbl$ voltage for 0.9 V is 710~$mV$ with a standard deviation of 20~$mV$. For 1 V, the mean $Rbl$ voltage is 692~$mV$ with a standard deviation of 21~$mV$. For 1.1 V, the mean $Rbl$ voltage is 674~$mV$ with a standard deviation of 19.2~$mV$.

\begin{figure*}[h]
\centering
\includegraphics[width = 0.88\textwidth]{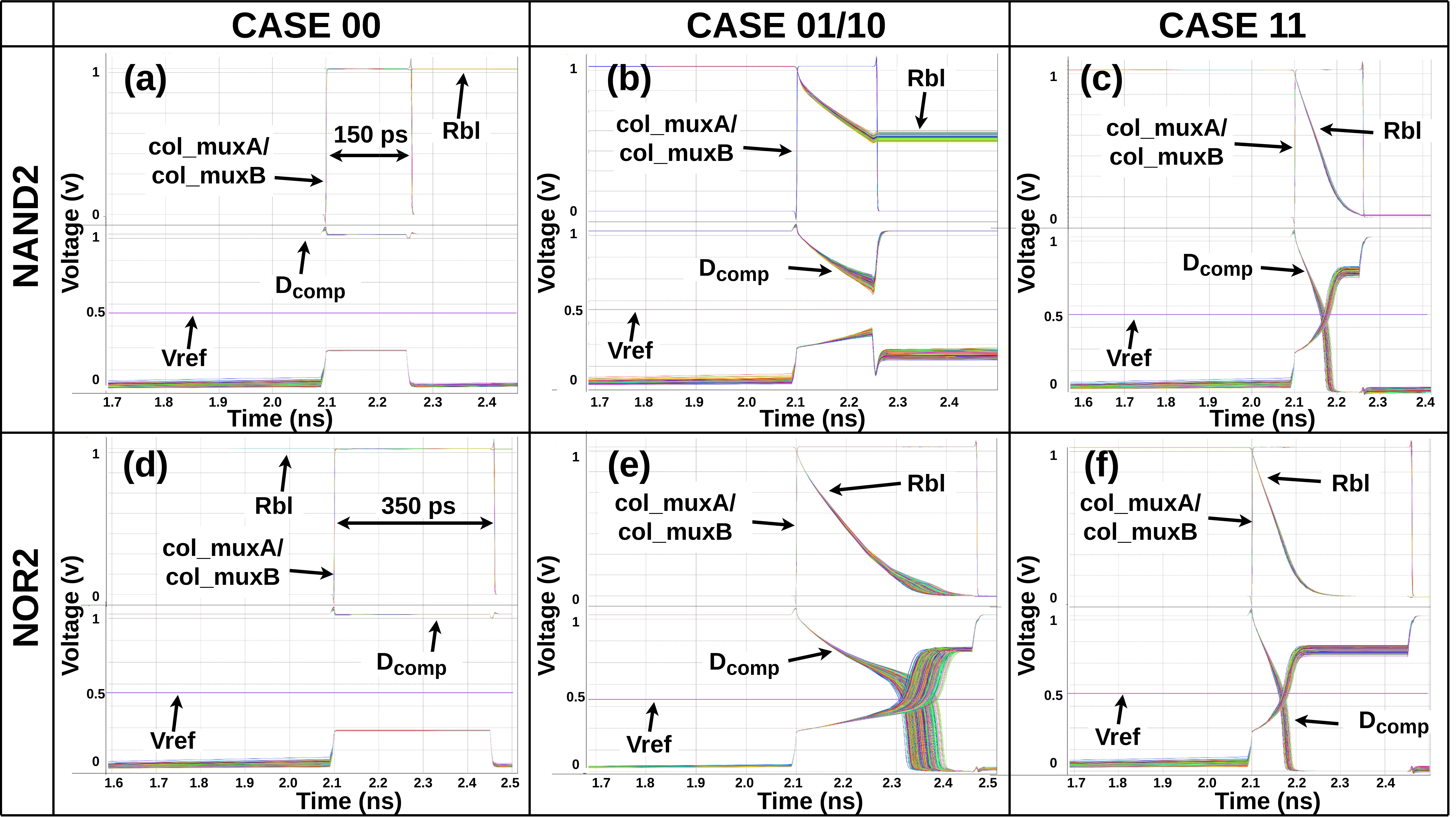}

\caption{Process variation analysis of the readout circuit considering all the cases for NAND2 and NOR2 operations show successful computational results of the sense amplifier considering 5000 samples with 3$\sigma$ deviations of $\pm$10\% length variation.}
\label{fig:sa_monte_carlo}

\end{figure*}
Figure~\ref{fig:sa_monte_carlo} demonstrates the robustness of the readout circuitry. We simulated 5000 samples with $\pm$10\% length variation and 3$\sigma$ deviations in the SA, shown in Figure~\ref{fig:novel_circuit}, considering an 8 KB SRAM rCiM architecture. Figures~\ref{fig:sa_monte_carlo} (a) and \ref{fig:sa_monte_carlo} (d) show the input case ``00'' for NAND2 and NOR2 operations, respectively. As $Rbl$ does not discharge in the ``00'' case, the output of the SA ($D_{comp}$) remains at logic ``1.'' For Figures~\ref{fig:sa_monte_carlo} (c), \ref{fig:sa_monte_carlo} (e) and \ref{fig:sa_monte_carlo} (f), corresponding to NAND2 input case ``11'' and NOR2 input cases ``01/10'' and ``11,'' the $Rbl$ completely discharges, resulting in a logic ``0'' for $D_{comp}$ value. In the NAND2 ``01/10'' case (Figure~\ref{fig:sa_monte_carlo} (b)), where the $Rbl$ partially discharges, the pulse width characterization ensures that $Rbl$ voltages do not drop below $Vref$ voltage, resulting in the correct $D_{comp}$ value of logic ``0.''

\subsection{Architecture Comparison with Previous Works}

\begin{table}[h]
\captionsetup{justification=centering} 
\caption{Comparison of the proposed rCiM architecture using 3 SRAM topologies with previous works show $2.6\times$ higher throughput and $1.6\times$ greater energy efficiency compared to~\cite{Wang_cim:19}, and achieving $2.12\times$ higher energy efficiency than~\cite{dac_20}.}
\label{tab:comparison}

\centering
\resizebox{\linewidth}{!}{
\renewcommand{\arraystretch}{1.5} 
\large 
\begin{tabular}{|>{\centering\arraybackslash}p{2.2cm}|*{9}{>{\centering\arraybackslash}p{2.3cm}|}}
\hline
 & \multicolumn{3}{c|}{\textbf{This work}} & \textbf{TVLSI'21~\cite{tvlsi_21}} & \textbf{ISSCC'19~\cite{Wang_cim:19}} & \textbf{DAC'20~\cite{dac_20}} & \textbf{DAC'19~\cite{dac_19}} & \textbf{TVLSI'23~\cite{Wang_cim:23}} & \textbf{JSSC'23~\cite{jssc_23}} \\
\hline
\textbf{Technology} & \multicolumn{3}{c|}{28nm} & 40nm & 28nm & 28nm & 28nm & 28nm & 28nm \\
\hline
\textbf{Cell Type} & \multicolumn{3}{c|}{10T dual read port} & 7T & 8T & 6T & 6T & 6T & 8T\\
\hline
\textbf{Array Size} & (256x256)x1 & (256x256)x3 & (512x256)x3 & 1Kb & (128x256)x4 & (128x128)x4 & 256x64 & 128x128 & 128x128\\
\hline
\textbf{Supply Voltage (V)} & \multicolumn{3}{c|}{1V} & 0.9 & 0.6-1.1 & 0.6-1.1 & 1 & 0.8 & 0.75\\
\hline
\textbf{Frequency (GHz)} & \multicolumn{3}{c|}{1GHz} & 0.1 & 0.475 & 2.25 & 2.2 & 0.633 & 0.113\\
\hline
\textbf{Throughput (GOPS)} & 88.2-106.6 & 264.83-320 & 529.66-640 & \parbox{2.3cm}{5.594\\44.752 \\(normalized\\ to 8KB)} & 32.7 & NA & \parbox{2.3cm}{560\\(normalized\\ to 8KB)} & \parbox{2.3cm}{162\\(normalized\\ to 8KB)} & 1851 \\
\hline
\textbf{Energy Efficient (TOPS/W)} & 8.64-10.45 & 8.64-10.45 & 17.18-20.77 & \parbox{2.3cm}{7.66\\8.86\\(normalized\\ to 28nm)} & 0.55 (mult), 5.27 (add) & 0.68 (mult), 8.09 (add) & NA & NA & 270.5\\
\hline
\textbf{Compute Density (GOPS/mm²)} &  \multicolumn{3}{c|}{551.25-666.25}  & 27 & 27.3 & NA & NA & NA & NA\\
\hline
\textbf{Type of Functions} & \multicolumn{3}{c|}{SRAM/ LOGIC (NAND, NOR, NOT)} & SRAM / NAND / NOR / XOR & Logic/ ADD/ SUB/ MULT/ DIV/ FP & SRAM/ LOGIC/ ADD/ MULT & SRAM/ Logic/ ADD/ Shift/ Copy & SRAM/ Logic/ ADD/ Compare  & SRAM/ Logic/ Copy/ Matrix Transpose\\
\hline
\end{tabular}%
}
\end{table}

A comparison of the proposed rCiM architecture with existing CiM architectures is presented in Table~\ref{tab:comparison}. The proposed architecture consumes 65~{$fJ$} per NAND2 operation and 116~{$fJ$} per NOR2 operation, achieving a throughput ranging from 88.2~{$GOPS$} to 106.6~{$GOPS$}, depending on NAND2 and NOR2 operations, with an 8 KB single macro implementation. The energy efficiency remains constant when transitioning from a single-macro to a three-macro implementation. While throughput increases by $3\times$ due to more operations being performed, the power consumption per cycle also increases by $3\times$, resulting in no net improvement in energy efficiency. However, when the array size is increased for the three-macro implementation, the power consumed by the computational circuits rises, but the control circuitry's overhead remains constant. This results in improved energy efficiency, as the increased throughput is greater than the increase in power consumption, leading to a higher overall energy efficiency. The proposed architecture achieves 551.25~{$GOPS/\text{mm}^{2}$} to 666.25~{$GOPS/\text{mm}^{2}$}, depending on the number of NAND2 and NOR2 operations. All throughput values of the compared works have been normalized to an 8 KB memory size.


Researchers in~\cite{tvlsi_21} propose a 7T bitcell and 2T switch are used for single-bit Boolean logic, addition, and multiplication operations. As this work is implemented in 40nm technology, we have used Dennard's power scaling law~\cite{Dennard:1974} to scale the power and obtain the energy efficiency. The proposed rCiM architecture achieves a $10\times$ higher frequency and 15\% greater energy efficiency with an 8 KB single macro implementation and a $2.2\times$ higher energy efficiency with a 16 KB three-macro implementation.

In~\cite{Wang_cim:19}, the transposable 8T cell performs multi-bit ``add" and ``multiplication" operations but has a lower frequency that results in higher energy/operation consumption. The proposed single-macro 8 KB rCiM architecture achieves $2.1\times$ higher frequency, resulting in an increase of throughput by $2.6\times$ and an increase in energy efficiency by $1.6\times$ when compared to~\cite{Wang_cim:19}.

In~\cite{dac_20}, the architecture boosts the bitline for computing to avoid read disturb issues, resulting in higher energy consumption. The proposed architecture overcomes read-disturb issues with a dedicated dual read-port bitcell, achieving $2.12\times$ higher energy efficiency with a 16 KB three-macro implementation compared to~\cite{dac_20}.

In~\cite{dac_19}, the authors present a high-speed 6T SRAM cell capable of performing bitwise addition, shift, and copy operations while mitigating read disturbance issues by incorporating an additional inverter and transistor to each bitline. Similarly, \cite{Wang_cim:23} introduces a 6T compute-SRAM architecture with dual-split-VDD assist in addressing read disturbance concerns. In contrast, our work utilizes dedicated read ports to eliminate read disturbance problems, which are prevalent in 6T SRAM-based CiM architectures. The throughput reported for both~\cite{dac_19} and~\cite{Wang_cim:23} is normalized to an 8 KB SRAM array. While these works demonstrate higher throughput than the single-macro implementation, they do not account for the additional write-back cycle required for output storage, which adds additional latency to each computation cycle. In~\cite{jssc_23}, the architecture stores the computation outputs directly in the same bitcell where the inputs are applied, resulting in significant latency and power savings. However, the reported throughput does not account for the additional latency required to read the operands and apply them as inputs to the bitcells. Additionally, designing this unconventional 8T SRAM requires a higher level of design expertise. In contrast, the proposed rCiM architecture operates at an $8.8\times$ higher frequency, leading to more efficient and conventional read and write operations.


%% file: conclusion.tex
\section{Conclusion}
\label{sec:conclusion}

This paper proposes an architectural exploration tool designed to identify the optimal rCiM cache topology tailored to specific logical operations. The novel rCiM architecture facilitates concurrent NAND2/NOR2/NOT operations using three-macro and six-macro topologies, significantly reducing latency for logical operations. Furthermore, the rCiM architecture incorporates a series resonance-based write driver, effectively lowering the consumed dynamic power during write operations by recycling the energy dissipated. The proposed algorithm utilizes only the RTL and a list of available SRAM topologies as input, streamlining the process of exploring the most energy-efficient topology for the given RTL. Comprehensive analysis conducted on EPFL combinational benchmark circuits demonstrates a notable average energy savings of 40.52\% across all the designs when employing the three-topology design implementations, as opposed to a single-macro implementation with the same macro size. 
The proposed three-topology implementation achieves $5.2\times$ higher throughput compared to~\cite{Lin_cim:21}, and $8.2\times$ higher throughput when compared with~\cite{Wang_cim:23}. The robustness analysis was conducted using Monte Carlo simulations with 5000 samples, considering temperature variations, $\pm10\%$ VDD, and $\pm10\%$ variations in transistor lengths. The analysis shows that the mean bitline discharge of 665~$mV$ with a standard deviation of 17~$mV$ for case ``$10/01$" of NAND2 operation, which falls within the sensing range of $VDD/2$ of the sense amplifier. Under the temperature and voltage variations the mean bitline discharge for case ``$10/01$" of NAND2 operation ranged between 710~$mV$ to 587~$mV$ with a standard deviation range of 27~$mV$ to 17~$mV$.